\documentclass[prd,aps,nofootinbib,floatfix,twocolumn,10 pt]{revtex4}

\usepackage{amssymb} \usepackage{amsmath,amssymb,graphicx,color,epsfig}
\usepackage{pstricks} \usepackage{float}

\begin{document}
\title{Geometrical measures of Non-Gaussianity generated from\\ single-field inflationary models}
\author{M.  Junaid$^{1,2}$\footnote{mjunaid@ualberta.ca}
and D.  Pogosyan$^{1}$\footnote{pogosyan@ualberta.ca}}
\affiliation{$^1$Department of Physics, University of Alberta, Edmonton,Canada,\\
$^2$National Centre for Physics, Islamabad, Pakistan.}

\date{\today}

\begin{abstract}
We calculate the third-order moments of scalar curvature
perturbations in configuration space for different inflationary models. We
develop a robust numerical technique to compute the bispectrum for different
models that have some features in the inflationary potential. From the
bispectrum we evaluate moments analytically in the slow-roll regime while we
devise a numerical mechanism to calculated these moments for non-slow-roll
single-field inflationary models with a standard kinetic term that are minimally
coupled to gravity. With the help of these third-order moments one can directly
predict many non-Gaussian and geometrical measures of cosmic microwave background distributions in the
configuration space. Thus, we devise a framework to calculate different
third-order moments and geometrical measures, e.g. Minkowski functionals
or the skeleton statistic, generated by different single-field models of inflation.
\end{abstract}

\maketitle

%%%%%%%%%%%%%%%%%%%%%%%%%%%%%%%%%%%%%%%%%%%%%%%%%%%%%%%
%			To Do List
%			Genus Formula correction
%			f_NL = k(\eta+2\eps) formmula correction in moments section
%			Numerical Analysis dlnPk/dlnk = -2eps -eta
%			Update picture "moments_logks.pdf" 
%			Genus Formula correction
%			v_f = v +S3 correction S_3
%			
%			
%			
%%%%%%%%%%%%%%%%%%%%%%%%%%%%%%%%%%%%%%%%%%%%%%%%%%%%%%%

\section{Introduction}\label{intro}
Recent results from ground- and balloon-based experiments as well as from 
WMAP and the Planck satellite have described the temperature anisotropies in 
the cosmic microwave background(CMB) with high precision\cite{boom,maxi,cbi,acbar,wmap1,planck1}. 
This primordial CMB is close to being Gaussian however there are some features 
and anomalies that are unexplained as shown in Ref. \cite{planck4} and references therein.
The study of non-Gaussian contributions in the cosmological perturbations is an ideal tool
to study the inflationary dynamics. Among the most direct data sets for studies
of inflation and non-Gaussianity are the CMB maps\cite{planck1,planck4,planck2,planck3}.

Inflation is the initial accelerated expansion of the early Universe. Inflationary expansion by
more than $65$ $e$-folds is needed to explain the observed homogeneity and isotropy of the Universe.
It is widely accepted that inflation is driven by the potential energy of a scalar inflaton
field slowly rolling down the potential. Inflation also successfully describes the creation 
of small inhomogeneities, needed to seed observed structure in the Universe,
as having been generated by quantum fluctuations in the inflaton field. The quantum fluctuations
also got stretched and became imprinted on CMB maps and other observables on
cosmological scales. Thus, inflation is able to explain not only why the Universe is so 
homogeneous and isotropic but also the origin of the structures in the Universe
\cite{Guth1,Guth2,Linde3,mukhanov2,Bardeen2,starobinsky83,mukhanov}.
Alternatives to inflation have been proposed but no other scenario is as simple and elegant as
inflation produced by scalar field(s)\cite{Linde1,Linde2,Kofman02}.

In single-field inflationary models generated inhomogeneities
are described via a single scalar perturbation field $\zeta(\mathbf{x})$. 
The statistical properties of $\zeta(\mathbf{x})$ are the main observable signatures
to distinguish inflationary models.
We know that if the perturbations are exactly Gaussian then all odd $n$-point
correlations functions vanish while all even $n$-point functions are related to
the two-point function. Thus, in the momentum space the Gaussian field $\zeta$
is completely described by the power spectrum $P_{\zeta}(k)$,
\begin{equation}
\left\langle{\zeta}_{\bold{k}}{\zeta}_{\bold{k}'}\right\rangle=
(2\pi)^3\delta^3(\bold{k}+\bold{k}')\frac{2\pi^2}{k^3}P_{\zeta}(k).
\end{equation}
Inflation generally predicts that the power spectrum is nearly flat
$P_\zeta(k) \propto k^{n_s-1} \approx \text{const}$ \cite{mukhanov2,starobinsky83,Bardeen2}.
The fact that the observed scalar spectral index
$n_s=0.9603\pm0.0073$ \cite{planck2} is close to
but not exactly unity is also considered by many to support the existence of inflation in
the early Universe. However, many different kinds of inflationary
models can be made compatible with observations of the power spectrum.
Thus the study of the non-Gaussian signatures is important to reduce the
degeneracy in inflationary models. Such signatures are contained in nontrivial
higher-order correlations starting with the cubic ones.

%For general set of initial perturbations, the higher
%order correlation functions give us more insight about the physics of early
%Universe. 
Similar to the power spectrum, for three-point correlations one can calculate the bispectrum
$B_{\zeta}(k_1,k_2,k_3)$ as a measure of the non-Gaussianity of the initial perturbations 
\begin{equation}
\left\langle\zeta_{\bold{k}_1}\zeta_{\bold{k}_2}\zeta_{\bold{k}_3}\right\rangle=
(2\pi)^3\delta^3(\bold{k}_1+\bold{k}_2+\bold{k}_3)B_{\zeta}(k_1,k_2,k_3)~.
\label{bis}
\end{equation}
The bispectrum carries much more information than the power spectrum as it contains three different length scales. 
It is thought that for basic single-field slow-roll inflation with a standard kinetic term the non-Gaussian effects
are small \cite{maldacena}, there are many models of inflation that give large, potentially detectable,
non-Gaussianity. Comparing non-Gaussian predictions with observations will help us
to constrain or rule out different inflationary models and give us more insight about the
physics of the early Universe.

In this paper we will focus on the study of non-Gaussianity through geometrical
measures that describe visual properties of the initial perturbations viewed as
a random field. Examples of standard measures of random fields are Minkowski
functionals, extrema statistics and also more novel measures such as skeleton
statistics. The simplest Minkowski functional is the Euler characteristic or genus
density of excursion sets as a function of threshold. It has been shown that for
mildly non-Gaussian field such local geometrical characteristics can be
expressed as a series of higher order moments to the perturbation field and
its derivatives \cite{Matsubara1,Bardeen3,PGP09,PPG11,Pogosyan1,Pogosyan2}. 
In particular for the Euler characteristic \cite{Matsubara1,PGP09}
\begin{eqnarray}
\chi_\textsc{3D}(\nu) \approx \left(\frac{\sigma_1}{\sqrt{3} \sigma} \right)^3
\frac{e^{-\frac{\nu^2}{2}}}{(2\pi)^{2}} 
\Bigg[ H_2(\nu) +\sigma \left(\frac{
\left\langle\zeta^3 \right\rangle }{6\sigma^4} H_5(\nu) \right. \Bigg.\notag \\
\Bigg.
\left. -\frac{3\left\langle\zeta^2\Delta\zeta \right\rangle}
{4\sigma^2\sigma_1^2} H_3(\nu)
- \frac{9\left\langle(\nabla\zeta)^2\Delta\zeta \right\rangle}{4\sigma_1^4}
H_1(\nu)
\right) \Bigg]\quad
\label{genus}
\end{eqnarray}
In the above expression $\sigma^2=\left\langle\zeta^2\right\rangle$,
$\sigma_1^2=\left\langle(\nabla\zeta)^2\right\rangle$, $\nu$ is the threshold in units of $\sigma$ and
$H_i(\nu)$ are Hermite polynomials. The first term in the expansion denotes the Gaussian part that is proportional
to $H_2(\nu)$ while other terms represent the first non-Gaussian correction in $\sigma$.

In this paper we have develop a robust mechanism to compute the third-order
moments such as $\left\langle \zeta^3\right\rangle$, $\left\langle \zeta^2\Delta\zeta \right\rangle$ and
$\left\langle (\nabla\zeta)^2\Delta\zeta \right\rangle$ for single-field models of
inflation. This links the non-Gaussianity generated by inflation to the
geometrical observables such as Euler characteristics and Minkowski functionals.

This paper is organized into six sections. In Sec.~II, we review the
theoretical framework of inflationary cosmology whereafter we will describe
the calculation of the three-point correlation function in momentum space and
the calculation of different third-order moments in configuration space. In Sec.~III, 
we will present our numerical technique for the calculation of three-point
function and briefly discuss different single-field models of inflation with
some features in the inflationary potential. In Sec.~IV, we will present
the calculation of moments in configuration space while in
Sec.~V we present the geometrical Minkowski functionals.
In the last section we will summarize our results and conclude with plans for
future work.

%%%%%%%%%%%%%%%%%%%%%%%%%%%%%%%%%%%%%%%%%%%%%%%%%%%%%%%%%%%%%%%%%%%%%%%%%%%%%%%%%%%%%%%%%%%%%%%%%%%%%%%%%%%% Theoretical Framework %%%%%%%%%%%%%%%%%%%%%%%%%%%%%%%%%%%%%%%%%%%%%%%%%%%%%%%%%%%%%%%%%%%%%%%%%%%%%%%%%%%%%%%%%%%%%%%%%%%%%%%%%%%%%%%%%%%%%%%%

\section{Theoretical Framework}
single-field inflation driven by a scalar field
$\phi$ is described by the following action in units of ($M_{pl}^{-2} = 8\pi
G=1$, $c=\hbar=1$)
\begin{eqnarray} S = \int d^4x \sqrt{-g} 
\left(
\frac{1}{2} R - \frac{1}{2} g^{\mu\nu} \partial_{\mu}\phi \partial_{\nu}\phi -V(\phi)
\right)
\label{action}
\end{eqnarray}
where $V(\phi)$ is the potential for the
inflaton field. In Friedmann cosmology with homogeneous and isotropic
background, the Friedmann equation for scale factor and Kline-Gordon equation
for inflaton field are given by
\begin{eqnarray}
H^2=\frac{1}{3}\left(\frac{1}{2}\dot{\phi}^2 +V(\phi) \right)\notag\\
\ddot{\phi}+3H\dot{\phi}+V_{,\phi}=0~.
\label{bg}
\end{eqnarray}
One can define the following slow roll parameters\footnote{Our definition of
$\eta=\frac{\dot{\epsilon}}{\epsilon H}=2\epsilon-2\eta_\textsc{H}$ is commonly
used in the studies of non-Gaussianity whereas
$\eta_\textsc{H}=-\ddot{\phi}/(\dot{\phi}H)$ is the Hubble slow roll parameter
used more commonly in the studies of inflation.} and the corresponding slow
roll conditions as
\begin{eqnarray}
\epsilon = -\frac{\dot{H}}{H}\ll1\text{, }
\eta = \frac{\dot{\epsilon}}{\epsilon H}\ll1~.
\end{eqnarray}
These slow roll
conditions $\epsilon\ll1, \eta\ll1$ ensure that the inflaton field rolls slowly
down the potential and the Universe inflates for significantly long period.
These slow roll parameters depend on potential of the inflaton field and the
model of inflation. For standard single-field inflation with quadratic
potential $V(\phi)=\frac{1}{2}m^2\phi^2$ these slow roll parameters are of
order $O(0.01)$ for inflaton field values $\phi>10M_p$.

\subsection{Calculation of 2-point and 3-point Functions in Momentum Space}
In this sections we will present the steps laid down by Maldacena to calculate
the two-point and the three-point correlation function of the scalar
perturbations \cite{maldacena}.
Firstly, one writes the action for the inflaton field given
in Eq.~\ref{action} using the Arnowitt-Deser-Misner(ADM) formalism. Secondly,
one expands the action to second order in perturbation theory for calculation
of two-point function and to third-order for the calculation of three-point
function. Thirdly, one quantizes the perturbations and imposes canonical
commutation relations. Next, one can define the vacuum state by matching the
mode function to Minkowski vacuum when the mode is deep inside the horizon that
fixes the mode function completely. Following these steps one can find the
power spectrum and the Bi-spectrum for scalar perturbations
\cite{maldacena,baumann1}.

In the ADM formalism the space-time is sliced into three-dimensional
hyper-surfaces $\Sigma$, with three metric $g_{ij}$, at constant time. The line
element of the space-time is given by
\begin{eqnarray}
ds^2=-N^2dt^2+g_{ij}(dx^i+N^idt)(dx^j+N^jdt)
\end{eqnarray}
where $N$ and $N^i$ and lapse and shift functions. In single-field inflation, we only have one
physically independent scalar perturbation. Thus, we perturb the metric and
matter part of the action and use the gauge freedom to choose the comoving
gauge for the dynamical fields $\phi$ and $g_{ij}$
\begin{eqnarray}
\delta\phi = 0,& g_{ij}=a^2( e^{2\zeta}\delta_{ij}+t_{ij})~,
\label{gauge}
\end{eqnarray}
where $\zeta$ is the comoving curvature perturbation at constant density
hyper-surface $\delta\phi = 0$. In this gauge, the inflaton field is
unperturbed and all scalar degrees of freedom are parameterized by the metric
fluctuations $\zeta(t,x)$ while the tensor perturbations are parameterized by
$t_{ij}$, that is both traceless and orthogonal $\partial_it_{ij}=t^i_i=0$. The
conditions in Eq.~\ref{gauge} fixes the gauge completely at non zero momentum
\cite{maldacena}.  The shift and lapse functions are not dynamical variables in 
ADM formalism hence they can be derived from constraint equations in terms of $\zeta$. 
We shall only study scalar perturbations in this paper.

Linear perturbation results are obtained if one expand the action to second
order in perturbation field $\zeta$
\begin{equation}
S_{(2)}= \int d^4x   a^3
\epsilon \left( \dot{\zeta}^2 - a^{-2}(\partial \zeta)^2 \right)
\end{equation}
which gives the following equation of motion for scalar perturbations
$v=z\zeta$ is Fourier space
\begin{equation}
v''_{\bold{k}} + \left(k^2
-\frac{z''}{z}\right)v_{\bold{k}} =0.
\label{Mukha}
\end{equation}
where $z=a\dot{\phi}/H$ and momentum $k$ is in reduced Planck units $M_{pl}$.
This is known as the Mukhanov equation for scalar perturbations\cite{mukhanov3}.
Now, one can calculate the 2-point function and the power spectrum
\begin{eqnarray}
\left\langle\hat{\zeta}_{\bold{k}}\hat{\zeta}_{\bold{k}'}\right\rangle =
(2\pi)^3\delta^3(\bold{k}-\bold{k}')\frac{2\pi^2}{k^3}P_{\zeta}(k) \notag \\
P_{\zeta}(k)=\frac{k^3}{2\pi^2}|u_k|^2
\label{PowerS}
\end{eqnarray}
where $u_k=v_k/z$ are the Fourier coefficients of $\zeta(x)$, the curvature
perturbations.

To obtain next order results in perturbation theory and to calculate
non-Gaussianity, one expands the action to third-order in scalar perturbations
in comoving gauge\cite{maldacena,DSeery}. After several integrations by parts
and dropping the total derivatives one finds the following third-order action
is often quoted in the literature\cite{maldacena,DSeery}
\begin{eqnarray}
S_{(3)} & =&  \int d^4x  \bigg(a^3\epsilon^2\zeta\dot{\zeta}^2
+ a\epsilon^2\zeta(\partial\zeta)^2-2a\epsilon\dot{\zeta}(\partial_i\zeta)(\partial_i\chi)\bigg.\notag \\
&+& \frac{a^3}{2}\epsilon\dot{\eta}\zeta^2\dot{\zeta}
+\frac{1}{2a}\epsilon(\partial_i\zeta)(\partial_i\chi)\partial^2\chi
+\frac{1}{4a}\epsilon(\partial^2\zeta)(\partial^2\chi)^2\notag \\
&+& \left.
2f(\zeta)\frac{\delta L}{\delta\zeta} \right)
\quad\label{S3o}
\end{eqnarray}
where the last term is variation of quadratic action is given by
\begin{eqnarray}
\frac{\delta L}{\delta\zeta}
&=&a\left(\frac{d\partial^2\chi}{dt}+H\partial^2\chi-\epsilon\partial^2\zeta
\right)\text{, }\partial^2\chi=\epsilon a^2 \dot{\zeta}\notag \\
f(\zeta)&=&\frac{\eta}{4}\zeta^2+\text{terms with derivatives of }\zeta.
\end{eqnarray}

Now, the calculation of the three-point function from the above action involves
integration over the time variable. But, if we use the action in Eq.~\ref{S3o}
to calculate the three-point function, the integral over time does not converge
to the end of inflation as pointed out by \cite{Arroja}. However, it was shown
in\cite{Arroja,Horner1,DSeery2} that this action can be converted into an equivalent
form, that gives a convergent three-point function, by adding a total
derivative term that is given by
\begin{equation}
-\frac{d}{dt} \left(\frac{\eta}{2} \epsilon a^3 \zeta^2 \dot{\zeta} \right)
\label{total}
\end{equation}
which brings the action to the following form.
\begin{widetext}
\begin{eqnarray} S_{(3)} &=&
\int d^4x\bigg(a^3\epsilon(\epsilon-\eta)\zeta\dot{\zeta}^2
+a\epsilon^2\zeta(\partial\zeta)^2  -
\frac{a}{2}\epsilon\eta\zeta^2\partial^2\zeta -
2a\epsilon\dot{\zeta}(\partial_i\zeta)(\partial_i\chi)
+\frac{1}{2a}\epsilon\partial^2\chi(\partial_i\zeta)(\partial_i\chi)
+\frac{1}{4a}\epsilon(\partial^2\zeta)(\partial_i\chi)^2 \bigg. \label{S3}  \\
\bigg.&+& 2g(\zeta)\frac{\delta L}{\delta\zeta} \bigg),\quad
g(\zeta)=\zeta\dot{\zeta}/H+\frac{1}{4a^2H^2}\left[
-(\partial\zeta)^2+\partial^{-2}(\partial_i\partial_j(\partial_i\zeta\partial_j\zeta))
\right] + \frac{1}{4a^2H}\left[
-(\partial\zeta)(\partial\chi)+\partial^{-2}(\partial_i\partial_j(\partial_i\zeta\partial_j\chi))
\right]\notag
\end{eqnarray}
\end{widetext}
Here the last term can be
eliminated with a field redefinition $\zeta \rightarrow \zeta_n+g(\zeta)$
because $g(\zeta)$ is only a function of derivatives of scalar perturbations
$\zeta(t,x)$ that vanish outside the horizon. The above third-order action is
an exact result without any slow roll approximations thus it is even valid for
models that deviate from slow roll conditions. Another feature of this action
is that it contains only first two slow roll parameters $\epsilon$ and $\eta$
while it is independent of derivative terms such as $\eta'$.

Finally, to calculate the 3-point function in momentum space we move to the
interaction picture and write the Hamiltonian for the action in
Eq.~\ref{action} as
\begin{equation}
H(\zeta) = H_0(\zeta) + H_{int}(\zeta)
\end{equation}
where $H_0$ is the quadratic part of the Hamiltonian while
$H_{int}$ represents all higher order terms in perturbation
theory\cite{maldacena}.  The three-point function in the interaction vacuum at
some time $\tau$ near the end of inflation is given by
\begin{eqnarray}
\left\langle
\zeta_{\bold{k}_1}(\tau)\zeta_{\bold{k}_2}(\tau)\zeta_{\bold{k}_3}(\tau)
\right\rangle=\notag \\ -i\int_{\tau_o}^{\tau} d\tau' a\left\langle
\left[\zeta_{\bold{k}_1}(\tau)\zeta_{\bold{k}_2}(\tau)\zeta_{\bold{k}_3}(\tau)
,H_{int}(\tau')\right] \right\rangle
\label{Hint}
\end{eqnarray}
To calculate 3-point
function, the  interaction Hamiltonian $H_{int}$ is just equal to $-S_{(3)}$
without the integral over time coordinate since the conjugate momenta vanish in
the ADM formalism.  Next, we quantize the perturbation field $\zeta(x)$ and
define the vacuum state. For this we expand the $\zeta(x)$ field into creation
and annihilation operators and use commutation relations of scalar field
$\left[\hat{a}_k,\hat{a}_{k'}^\dagger \right]=\delta_{kk'}$ to get the
following result
\begin{eqnarray}
\left\langle
\zeta_{\bold{k}_1}\zeta_{\bold{k}_2}\zeta_{\bold{k}_3} \right\rangle = i
(2\pi)^3\int_{-\infty}^{\tau_{end}} d\tau \left(-2a^2\epsilon^2 u^*_1 u'^*_2
u'^*_3 \frac{\bold{k}_1.\bold{k}_2}{k_2^2} \right. \notag \\
\left.+2a^2\epsilon(\epsilon-\eta) u^*_1 u'^*_2 u'^*_3
-a^2\epsilon(2\epsilon\bold{k}_1.\bold{k}_2+\eta k_3^2) u^*_1 u^*_2 u^*_3
\right.\notag \\ + \frac{a^2}{2}\epsilon^3 u^*_1 u'^*_2 u'^*_3 k_1^2
\frac{\bold{k}_2.\bold{k}_3}{k_2^2 k_3^2 } +\frac{a^2}{2}\epsilon^3u^*_1 u'^*_2
u'^*_3 \frac{\bold{k}_1.\bold{k}_2}{k_2^2}+c.c.\notag \\ \bigg.+ \text{distinct
permutations}\bigg) \left.\prod_{i=1}^3 u_{i}(\tau_{end})\right.\delta^3(\sum_j
\bold{k}_j)\quad \label{3pt}
\end{eqnarray}
The choice of the vacuum is
specified by the choice of mode function $u_k$ selection. This is the main
formula for the three-point function. This expression will be used in the
sections to come for the exact numerical calculation of three-point function in
momentum space.

\subsection{Non-Gaussian parameters and $f_\textsc{NL}$}
The integral relation
for 3-point function given in Eq.~\ref{3pt} can be analytically evaluated in
the slow roll limit. Ignoring $\epsilon^3$ terms in Eq.~\ref{3pt}, gives us the
following result that was first derived by Maldacena\cite{maldacena}.
\begin{eqnarray}
\left\langle\zeta_{\bold{k}_1}\zeta_{\bold{k}_2}\zeta_{\bold{k}_3}\right\rangle
= (2\pi)^7\delta^3(\bold{k}_1+\bold{k}_2+\bold{k}_3)\frac{(P_k^{\zeta})^2}{\prod_ik_i^3} \mathcal{A}\label{3A} \\
\mathcal{A} =\frac{\eta^*-\epsilon^*}{8}\sum_i k_i^3 +
\frac{\epsilon^*}{8}\sum_{i \ne j} k_ik_j^2 + \frac{\epsilon^*}{K}\sum_{i>j}
k_i^2k_j^2\quad \label{A3p}
\end{eqnarray}
where $*$ denotes the $\epsilon$ and
$\eta$ values at horizon crossing. The quantity $\mathcal{A}$ is a convenient
measure of non-Gaussianity in the perturbation field. The relationship between
$\mathcal{A}$ and bispectrum is given by
\begin{equation}
B_{\zeta}(k_1,k_2,k_3)=(2\pi)^4\frac{(P_k^{\zeta})^2}{\prod_ik_i^3}
\mathcal{A}.
\end{equation}

The bispectrum and $\mathcal{A}$ are general measures of non-Gaussianity
however both these quantities are highly scale dependent. Thus, the three-point
correlation is often described in terms of a local dimensionless non-linearity
parameter $f_\textsc{NL}$. This non-linearity parameter $f_\textsc{NL}$ was
first introduced as a measure of local non-Gaussianity described by
\begin{eqnarray}
\zeta(x)=\zeta_L(x)-\frac{3}{5}f_\textsc{NL}\left(
\zeta_L(x)^2- \left\langle\zeta_L(x)^2\right\rangle\right)~,
\label{fnl}
\end{eqnarray}
where factor $\frac{3}{5}$ is a matter of convention. For local non-Gaussianity,
the quantity $\mathcal{A}$ is expressed via the non linearity parameter
$f_\textsc{NL}$ as
\begin{equation}
\mathcal{A}_{local}=-\frac{3}{10}f_\textsc{NL}\sum_i k_i^3~.
\label{anl}
\end{equation}
Beyond local model we can define a generalized $f_\textsc{NL}$
for general kind of non-Gaussianity by the following equation, that also has
the advantage of being nearly scale independent.
\begin{equation}
f_\textsc{NL}\equiv-\frac{10B_{\zeta}(k_1,k_2,k_3)\prod_i
k_i^3}{3(2\pi)^4(P_k^{\zeta})^2\left(\sum_i  k_i^3\right)}
\label{fg}
\end{equation}
In this paper we will be using this generalized definition of
$f_\textsc{NL}$ instead of $\mathcal{A}$ or bispectrum that are both scale
dependent quantities. The $f_\textsc{NL}$ depends on the shapes of three-point
function triangles and it will depend upon scale of triangles if there are
features in the inflaton potential. Over the recent years $f_\textsc{NL}$ has
become a widely used measure of non-Gaussianity\cite{Komatsu1}.

\subsection{Introduction to Moments}
Non-Gaussianity can also be studied
through the higher order moments of the perturbation field in configuration
space. Analysis of these moments provide a robust measures of non-Gaussianity
and has also become an important field of investigation\cite{wmap2,Park}. In
this paper we will study how different inflationary models can predict
different non-Gaussian and other geometrical parameters in configuration space.
These moments can provide important information on the geometrical properties
of the physical fields, e.g. CMB temperature fluctuations,
and give us non-Gaussianity observables such as
extrema counts, genus and skeleton\cite{Pogosyan1}. 

To calculate the third-order moments we have to take the inverse Fourier
transform of three-point function in momentum space. These third-order moments
also contain derivatives of the perturbation field $\partial_i\zeta$ and
$\Delta\zeta$. The complete and independent set of moments that are needed to
calculate the observables such Euler characteristic are given by
\begin{eqnarray}
\left\langle \zeta^3(x) \right\rangle
=\int \frac{dk_1^3 dk_2^3 dk_3^3}{(2\pi)^9}
\left\langle\zeta_{\bold{k}_1}\zeta_{\bold{k}_2}\zeta_{\bold{k}_3}\right\rangle
e^{i(\bold{k}_1+\bold{k}_2+\bold{k}_3).\bold{x}}\label{Mm1}\\ \left\langle
\zeta^2(x)\Delta\zeta(x) \right\rangle
= \int \frac{dk_1^3 dk_2^3
dk_3^3}{(2\pi)^9}\left\langle \zeta_{\bold{k}_1}\zeta_{\bold{k}_2}\zeta_{\bold{k}_3}
\right\rangle\notag \\
\times k_3^2 e^{i(\bold{k}_1+\bold{k}_2+\bold{k}_3).\bold{x}}\label{Mm2} \\
\left\langle(\nabla\zeta(x))^2\Delta\zeta(x) \right\rangle = \int \frac{dk_1^3 dk_2^3
dk_3^3}{(2\pi)^9}\left\langle \zeta_{\bold{k}_1}\zeta_{\bold{k}_2}\zeta_{\bold{k}_3}
\right\rangle\notag \\
\times \bold{k}_1.\bold{k}_2 k_3^2
e^{i(\bold{k}_1+\bold{k}_2+\bold{k}_3).\bold{x}}~.
\label{Mm3}
\end{eqnarray}
These moments are always scaled by the corresponding variances $\sigma$ and
$\sigma_1$ in all physical observables like Minkowski functionals and Euler
characteristics.

The details of calculating these moments for different inflationary models will
be presented in section \ref{MG}. However, one can the calculate these moments
for local non-Gaussianity given by Eq.~\ref{fnl} in configuration space and
there values are given by
\begin{eqnarray}
\frac{\left\langle \zeta^3(x)
\right\rangle_{local}}{\sigma^4} = -\frac{18}{5}f_\textsc{NL}\\
\frac{\left\langle \zeta^2(x)\Delta\zeta(x) \right\rangle_{local}}{\sigma^2\sigma_1^2} = \frac{24}{5}f_\textsc{NL}\\
\frac{\left\langle(\nabla\zeta(x))^2\Delta\zeta(x)
\right\rangle_{local}}{\sigma_1^4}=\frac{8}{5}f_\textsc{NL}
\end{eqnarray}
The above values of moments are well known results for local non-Gaussianity
that is independent of any model related study of inflation.

%%%%%%%%%%%%%%%%%%%%%%%%%%%%%%%%%%%%%%%%%%%%%%%%%%%%%%%%%%%%%%%%%%%%%%%%%%%%%%%%%%%%%%%%%%%%%%%%%%%%%%%%%%%% Numerical Technique %%%%%%%%%%%%%%%%%%%%%%%%%%%%%%%%%%%%%%%%%%%%%%%%%%%%%%%%%%%%%%%%%%%%%%%%%%%%%%%%%%%%%%%%%%%%%%%%%%%%%%%%%%%%%%%%%%%%%%%%%

\section{Numerical Technique} 
\subsection{Calculation of Mode Function and Power Spectrum} 
To calculate the power spectrum of scalar perturbations we need
to solve the background equations of motion Eqs.~\ref{bg} and the Mukhanov
equation Eq.~\ref{Mukha} written in terms of conformal time
\begin{eqnarray}
a'' =\frac{a}{6}\left( -\phi'^2 + 4a^2V(\phi) \right) \notag  \\
\phi''=-2\frac{a'}{a}\phi'-a^2V_{,\phi} \notag  \\ v''_k=-\left(
k^2-\frac{z''}{z} \right)v_k\text{, } z = a\sqrt{2\epsilon}
\end{eqnarray}
where primes $'$ donate derivatives with respect to conformal time. These are
coupled differential equation with first two representing the background and
last equation for scalar perturbations.  Numerically, it is more convenient to
work out the differential equation for $u_k$ rather then $v_k$ since we finally
require $u_k=v_k/z$ to calculate the non-Gaussianity and power spectrum. Thus,
we convert the Mukhanov equation to perturbation equation for $u_k$
\begin{eqnarray}
u''_k=- k^2 u_k -2\frac{z'}{z}u'_k.
\end{eqnarray}
We solved these equations for background and scalar perturbations using
the Runge-Kutta
method of order four.  The initial condition for solving these differential
equations are given by the following equations
\begin{eqnarray}
a(\tau_0)&=&1\text{,   }\phi(\tau_0)=\phi_0\\
v_k(\tau_0)&=&\sqrt{\frac{1}{2k}}\text{,   } v'_k(\tau_0)=-i\sqrt{\frac{k}{2}}.
\end{eqnarray}
We chose initial value of inflaton field $\phi_0$ such that
Universe expands for $70$ e-folds for quadratic potential. The mode functions
originate deep inside the horizon that correspond to, our choice of vacuum, the
Bunch-Davies vacuum.

The above equations of motion are for single-field inflation with standard
kinetic term with any potential $V(\phi)$. However in this paper we will
specifically study quadratic inflation $V(\phi)=\frac{1}{2}m^2\phi^2$ as our
base model to check our numerics. We will also study two other models that have
features added to the quadratic model that are known as the step and resonance
models, details of which are given in the next section. 

\begin{figure}[htbp]
\begin{center}
\includegraphics[height=6.3cm]{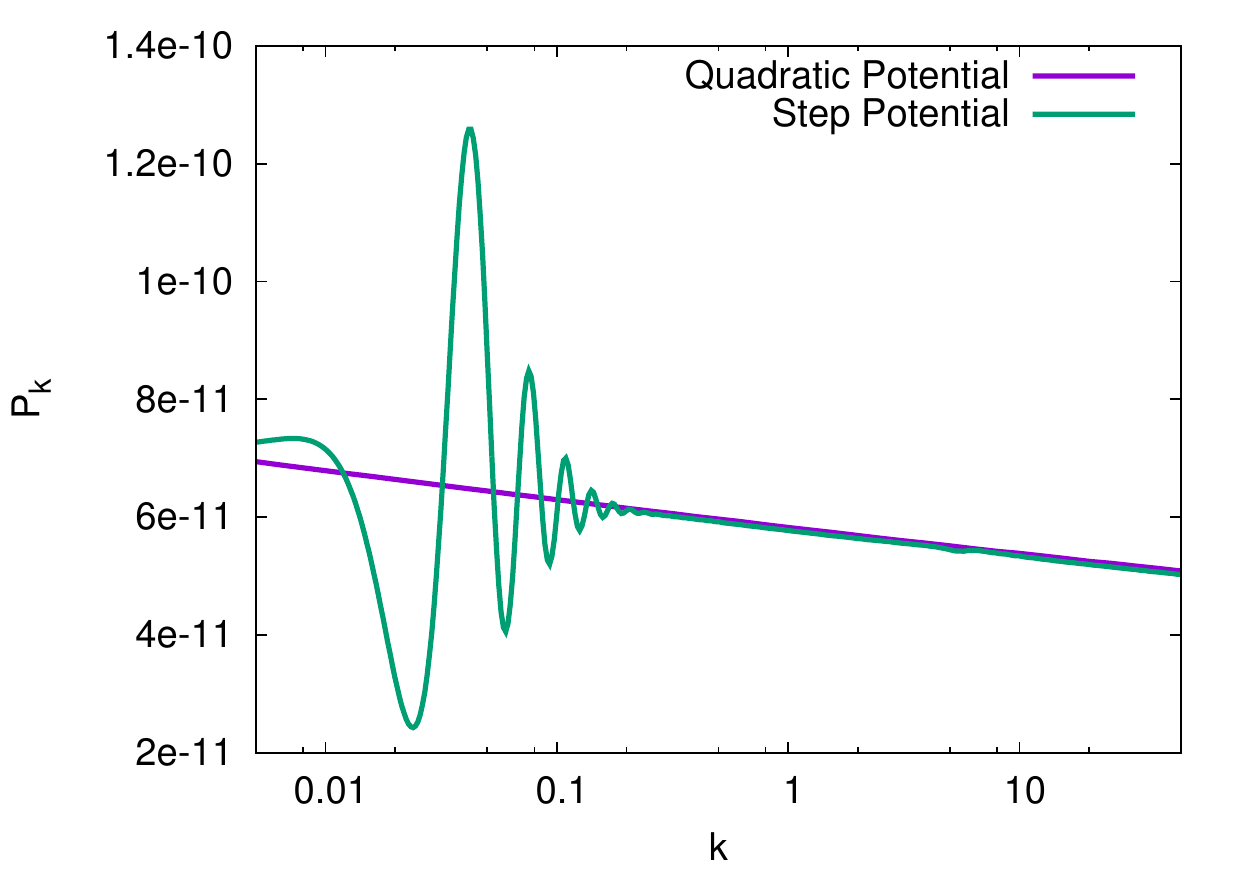}
\caption{Power spectrum of scalar perturbations $P_{\zeta}(k)$ against momenta
$k$ for quadratic potential and step potential with $c=0.002$, $d=0.02\;M_{pl}$ and
$\phi_s=15.86\;M_{pl}$. }
\label{Pk}
\end{center}
\end{figure}

In Fig \ref{Pk} we have presented the calculation of dimensionless power
spectrum according to Eq.~\ref{PowerS}. The power spectrum is mildly dependent
on $k$ for the quadratic potential with $\frac{d\ln P_{\zeta}}{d\ln
k}=-2\epsilon^*-\eta^*$. On the other hand for the step potential, due to the
breaking of the slow roll condition because of a sharp step in the potential,
we see oscillating power spectrum near the step but as we move away from the
step it follows the same behaviour of the quadratic potential (Fig. \ref{Pk}).

\subsection{three-point function and Cesaro Sum} After numerically solving the
background equations and the equation for scalar perturbations, we insert these
solutions back into Eq.~\ref{3pt} to calculate the three-point function in
momentum space. The three-point correlation function is numerically challenging
task as it involves integrations that arise from equation (\ref{3pt}). The
integrands consist of three factors of $u_k$ or $u'_k$ multiplied by the
background factors of $a$, $\epsilon$ and $\eta$. The scalar perturbation
function $u_k$ oscillates before horizon crossing at $\tau_0$, while after
horizon it freezes out. Thus, the integration consists of two parts, before
horizon crossing (BHC) part and after horizon crossing (AHC) part
\begin{eqnarray}
\int_{-\infty}^{\tau_{end}} d\tau I(\tau) =
\int_{-\infty}^{\tau_0} d\tau I(\tau)+\int_{\tau_0}^{\tau_{end}} d\tau I(\tau)
\label{integ}
\end{eqnarray}
where $\tau_0$ is the horizon crossing point of
the largest $k$ mode in the three-point correlation function and $I(\tau)$ is
the integrand of the 3-point function given in Eq.~\ref{3pt} that contains
background factors and product of three oscillating mode functions. The BHC and
AHC parts of integration present different numerical challenges as the first
has growing oscillations, as $\tau$ approaches negative infinity,  while for
the AHC part we have to regularize the three-point function by adding a total
derivative term(Eq.~\ref{total}) in the action. Without adding this term in the
action, the AHC part of the integral is divergent as one of the term
$a^3\epsilon\dot{\eta}\zeta^2\dot{\zeta}$ in the initial action grows as the
scale factor \cite{Arroja,Horner1}.

The contribution to the integral that arises from before horizon crossing poses significant technical
challenges. In conformal time the initial big bag singularity is pushed back in
conformal time to $\tau \rightarrow -\infty$. Thus, the scalar perturbations
start deep inside the horizon and keep oscillating till horizon crossing point
$\tau_0$ of the largest $k$ mode in the 3-point function. Now, there are
different methods to numerically evaluate an oscillating integral over an
infinite range. If we cutoff this infinite integral to some finite value, due
to large oscillations this induces an spurious contribution of $O(1)$.
Numerically it was shown that these kind of integrals can be evaluated by
introducing an arbitrary damping factor into the integrand but this damping
factor needs to be chosen carefully\cite{XChen1}. Other techniques, such as
boundary regularizaion, for evaluating such integrals are even more complex
\cite{XChen3,Horner1}.

We have developed a different numerical technique, which is numerically more
robust and elegant, using the Cesaro resummation of improper series. For
oscillating integrand $I(\tau)$, the following expressions gives the definition
of Cesaro integration
\begin{eqnarray}
\int_{-\infty}^{\tau_0} d\tau I(\tau)
\equiv \lim_{\tau \rightarrow -\infty} \frac{1}{\tau_0-\tau}
\int_{\tau_0}^{\tau} d\tau' \left( \int_{\tau_0}^{\tau'} d\tau'' I(\tau'')
\right)~.
\label{cesaro}
\end{eqnarray}
This gives a specific definition to the
improper integral on the left hand side whereas the right hand side is an
average over the partial integrals that give convergent result for a wide range
of improper integrals\cite{cesa}. However, we extended this method further and
we defined a higher order Cesaro integral, with one additional average, to
further improve the convergence as
\begin{eqnarray}
\lim_{\tau \rightarrow
-\infty} \frac{1}{\tau_0-\tau} \int_{\tau_0}^{\tau} d\tau'
\frac{1}{\tau_0-\tau'} \int_{\tau_0}^{\tau'} d\tau'' \left(
\int_{\tau_0}^{\tau''} d\tau''' I(\tau''') \right)~.
\label{cesaro2}
\end{eqnarray}
In our numerical program we have used this Cesaro integral with
additional average Eq.~(\ref{cesaro2}) for faster convergence.
\begin{figure}[htbp]
\begin{center}
\includegraphics[height=6.3cm]{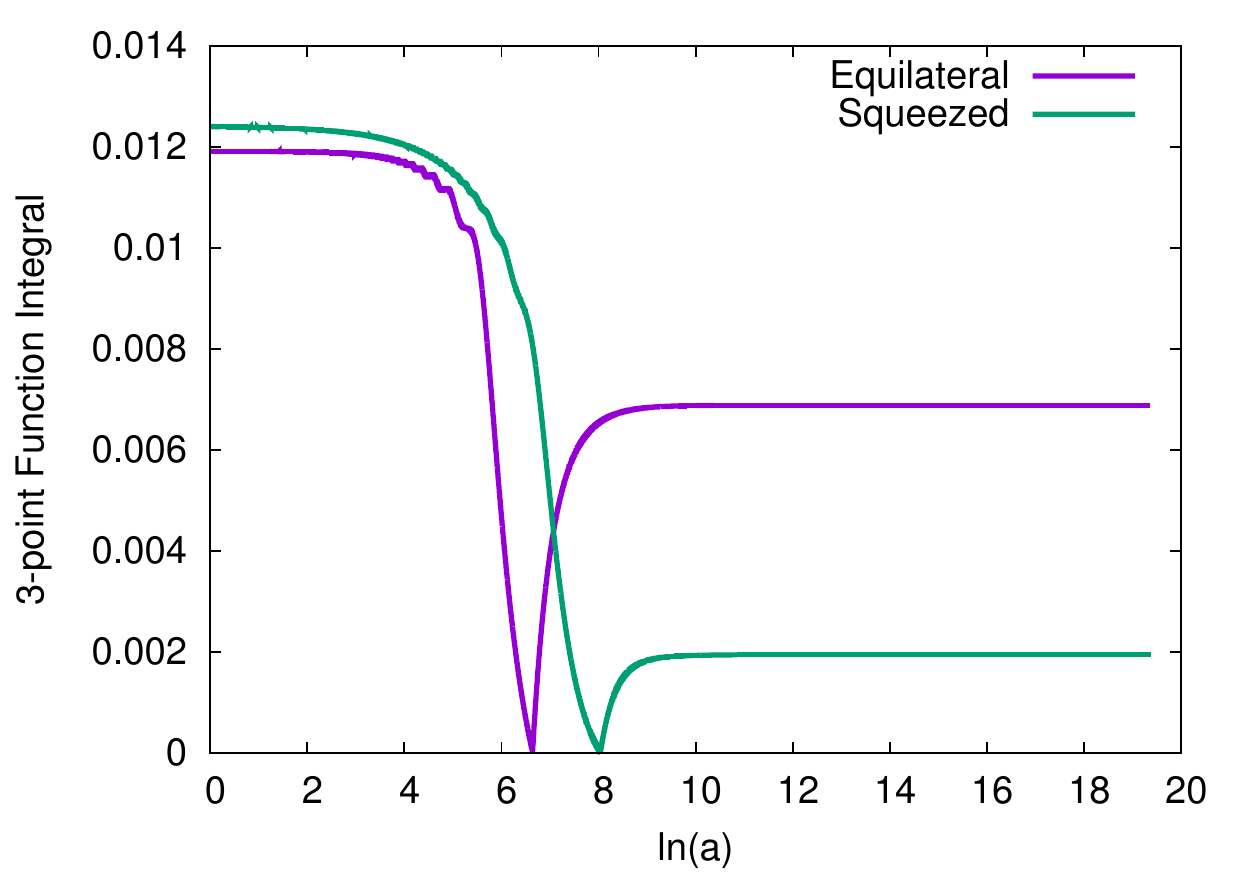}
\caption{three-point correlation function, or generalized
$f_\textsc{NL}$, integral plotted against the number of e-folds $N=\ln(a)$ for
Equilateral and Squeezed triangles with horizon crossings at $6.5$ and $8$
e-folds. The generalized $f_\textsc{NL}$ is just the sum of the two
asymptote(plateau) values at small and large e-folds corresponding to the two
integrals in Eq \ref{integ}.}\label{uuu}
\end{center}
\end{figure}

This method quickly gives convergent results without introducing any artificial
damping factors. This can be seen in Fig.~\ref{uuu} which shows the three-point
function integral result plotted against the number of e-folds for an
equilateral triangle and squeezed triangle cases. In this figure horizon
crossings occur $\tau_0$ that corresponds to e-folds values of $6.5$ and $8$
for equilateral and squeezed triangle. This Fig~\ref{uuu} describe two
different integration regimes BHC $\tau<\tau_0$ and AHC $\tau>\tau_0$. In BHC
regime, we integrate in backward direction from the horizon crossing points
using the Cesaro Integral Eq.~\ref{cesaro2}. Our technique converges very
quickly as can be seen that the integral plateaus as we go 5-6 e-folds before
horizon crossing points. In the AHC regime, we integrate in the forward
direction that also plateaus soon after horizon crossing. Thus, the three-point
function integral, or generalized $f_\textsc{NL}$, is just the sum the two
asymptote(plateau) values in the before and after horizon crossing regimes for
each kind of triangle.
 
\begin{figure}[tbp]
\begin{tabular}{c}
\includegraphics[height=6.3cm,clip=a]{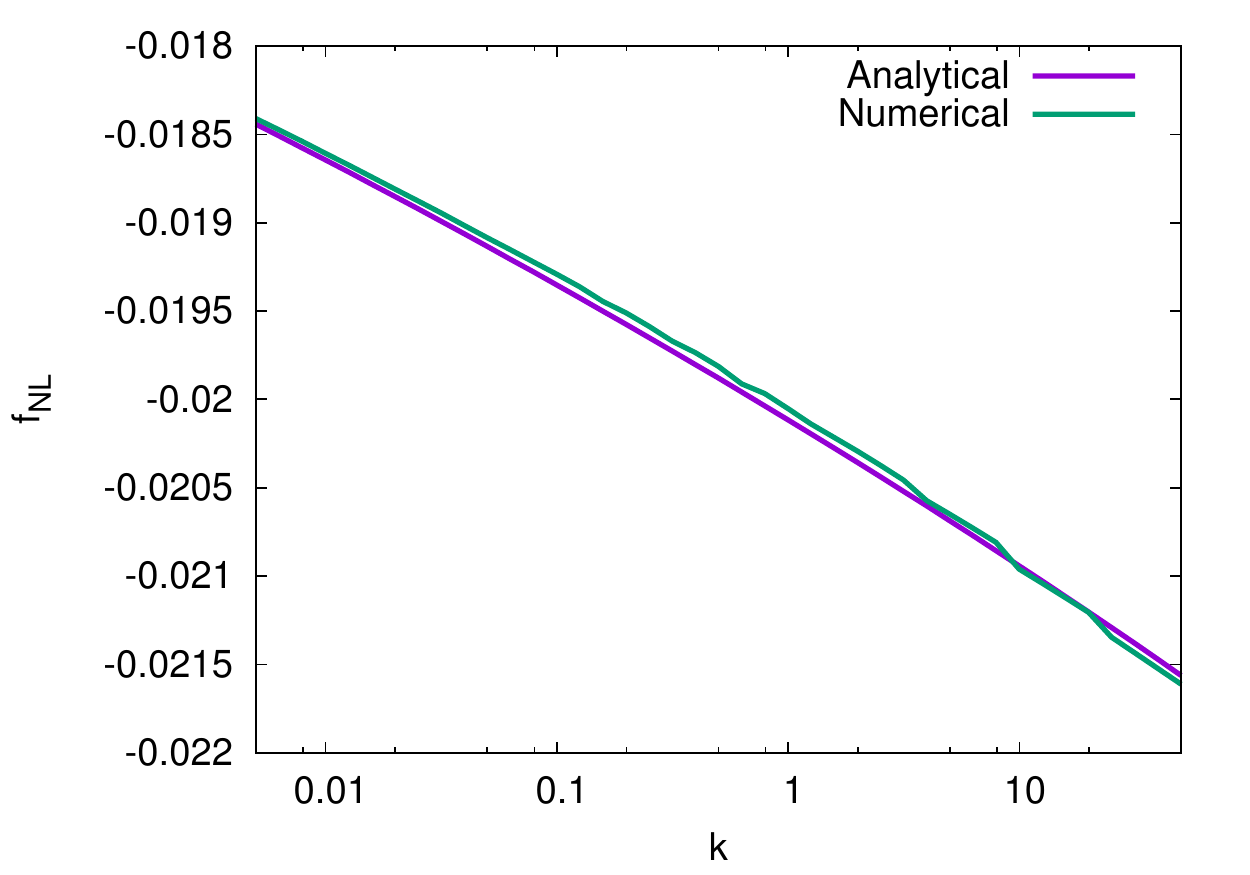}\\
\includegraphics[height=6.3cm,clip=b]{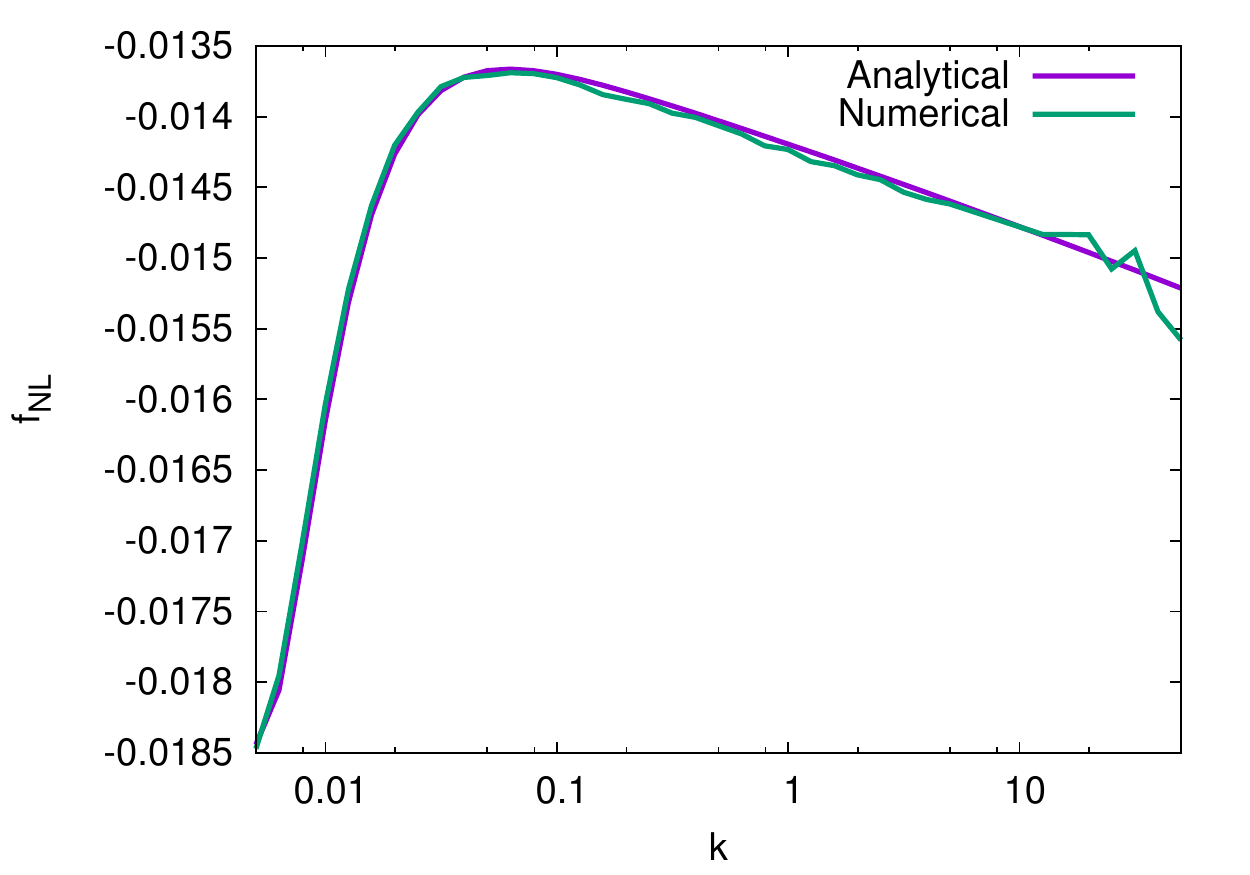}
\end{tabular}
\caption{Numerical calculation of $f_\textsc{NL}$ for an equilateral triangle case($k=k_1=k_2=k_3$)
alongside analytical result (Fig. a) for quadratic potential. Second plot show
results for a squeezed triangle $k_1=0.005\;M_{pl}$ while $k=k_2=k_3$ are on the
x-axis (Fig. b) for quadratic potential.}
\label{fnlq}
\end{figure}

\begin{figure}[tbp]
\includegraphics[height=6.3cm]{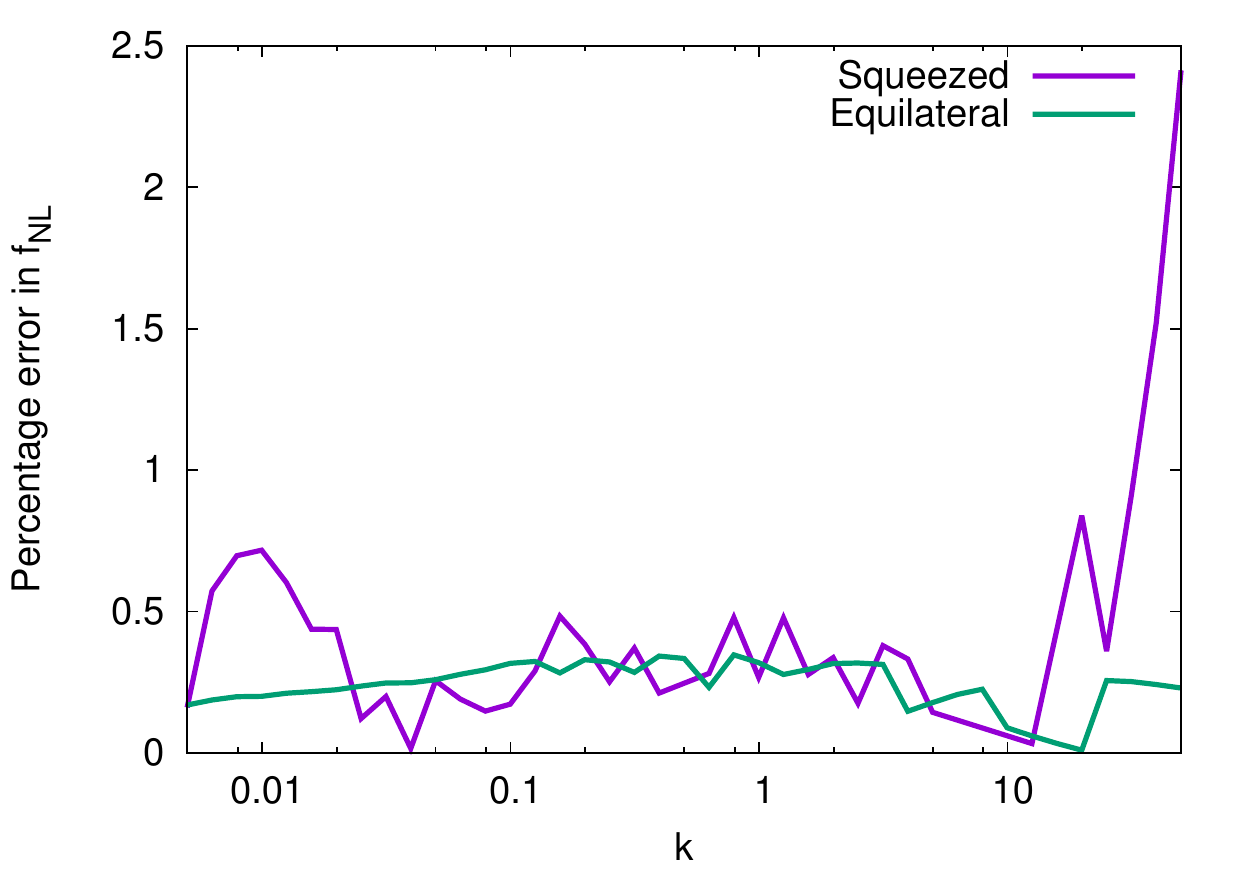}
\caption{Percentage error in calculation of $f_\textsc{NL}$ for an equilateral triangle case when
compared with analytical results plotted against $k=k_1=k_2=k_3$. While for squeezed triangle
percentage error is plotted against $k=k_2=k_3$ while fixed $k_1=0.005\;M_{pl}$ for quadratic potential. }
\label{error}
\end{figure}

To test our procedure, we have calculated the three-point function numerically
for $\frac{1}{2}m^2\phi^2$ potential and compared it with the corresponding
analytical results given by Eq.~\ref{A3p}. The numerical results when compared
the analytical results are plotted in Fig. \ref{fnlq}(a, b) for equilateral and
squeezed triangles. As can be seen that our numerics follows very closely the
analytical results for the quadratic potential. The Fig. \ref{error} shows that
our numerical technique is accurate to below one percent error for equilateral
and squeezed triangles. In the extreme squeezed triangle limit $k\approx5$ in
Fig.~\ref{error} the errors are somewhat large as a compromise was made in the
calculation due to the time and memory constraints.

\subsection{Application to Models with Significant Non-Gaussianity}
Single
field inflation with quadratic potential, $V(\phi)=\frac{1}{2}m^2\phi^2$, is
categorized as chaotic inflation. The dynamics of inflation can be solved
exactly in the slow roll limit for quadratic potential, thus we have used it as
our base model for our study. However, it is was shown by Maldacena that single
field slow roll models of inflation give rise to small non-Gaussianity of
$O(\epsilon)$ as can seen in Fig.~\ref{fnlq} and Eq.~\ref{A3p}. In this paper
we take the inflaton field mass to be $m=10^{-6}M_p$. 

In single-field models of inflation significant non-Gaussianity can arise from
different kinds of non trivial potentials, for example adding some features to
quadratic potential. One such model consist of localized breaking of slow roll
conditions. This can be achieved be adding a step in the quadratic potential as
proposed by \cite{Adams}
\begin{eqnarray}
V_{step}(\phi)=\frac{1}{2}m^2\phi^2\left ( 1+ c\tanh\frac{\phi-\phi_s}{d}
\right).
\label{step}
\end{eqnarray}
where $c$ is the height of the step and
$d$ the width of the step centred at $\phi_s$. This model has been used to
improve the fit between LCDM and observed power spectrum and we have used the
best fit parameters for the step potential\cite{Adams}. When inflation rolls
down through this step it goes through a sharp acceleration in inflationary
dynamics. The three-point correlation function is proportional to $\epsilon$
and $\eta$ parameters and deviation from slow roll gives rise to large
interaction of modes near the step. This gives rise to large non-Gaussianity
and generalized $f_\textsc{NL}\approx\frac{7c^{3/2}}{\epsilon d}$\cite{XChen1}.
It important to mention here that $\epsilon$ does not change much by the step
potential however $\eta\approx\frac{7c^{3/2}}{\epsilon d}$ gives dominant
contribution to the 3-point function. Now, the three-point function integral
Eq.~\ref{3pt} gets most of its contribution near the horizon crossing part of
the modes. Thus, due to the step in the potential, the modes that cross the
horizon near the step get a sharp kick that gives rise to large non-Gaussianity\cite{XChen3,Amjad}.
\begin{figure}[htbp]
\begin{center}
\includegraphics[height=6.3cm]{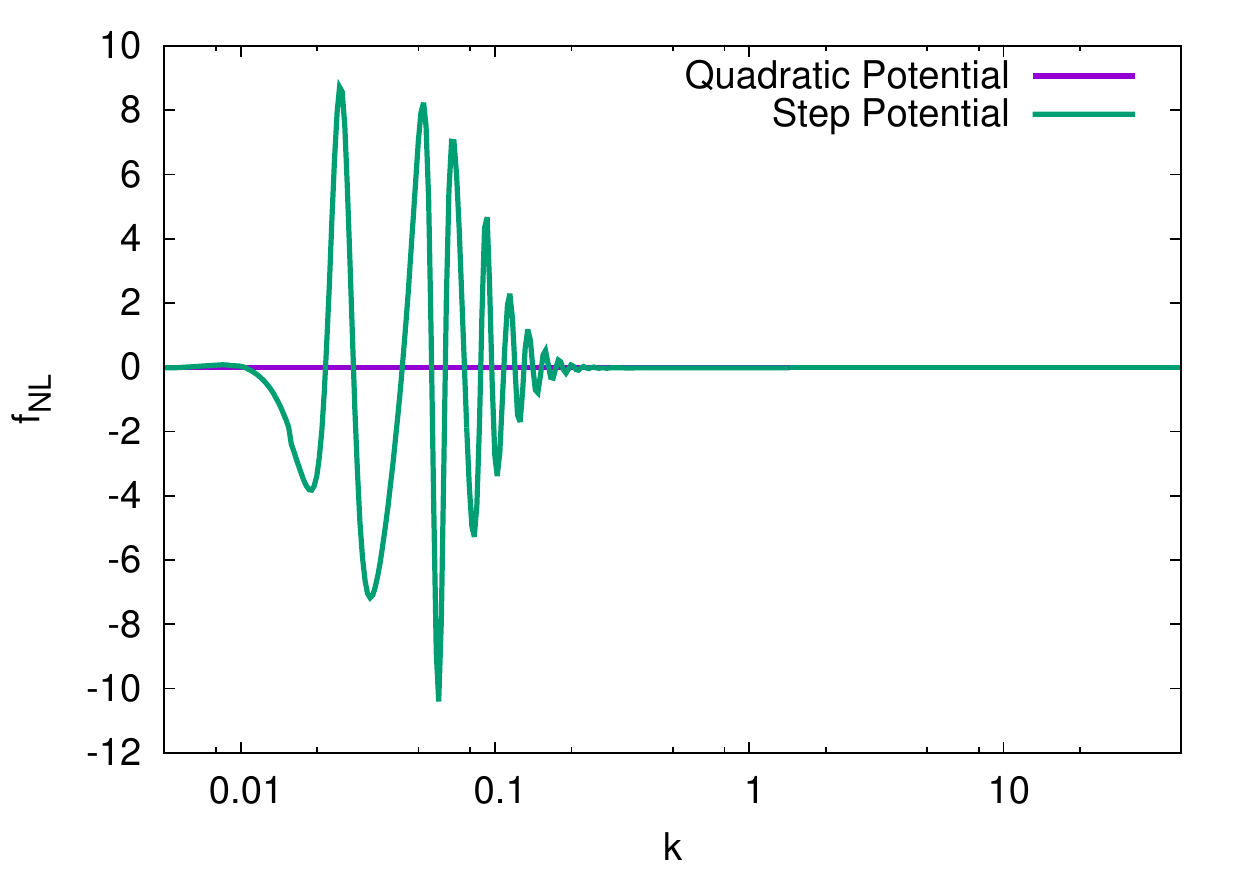}
\caption{$f_\textsc{NL}$ for the step potential model plotted against $k=k_1=k_2=k_3$ for 
equilateral triangle case with step height and width given by $c=0.002$ and
$d=0.02M_{pl}$ respectively while step is located at $\phi_s=15.86\;M_{pl}$.}
\label{fnls}
\end{center}
\end{figure}
\begin{figure}[htbp]
\begin{center}
\includegraphics[height=6.3cm]{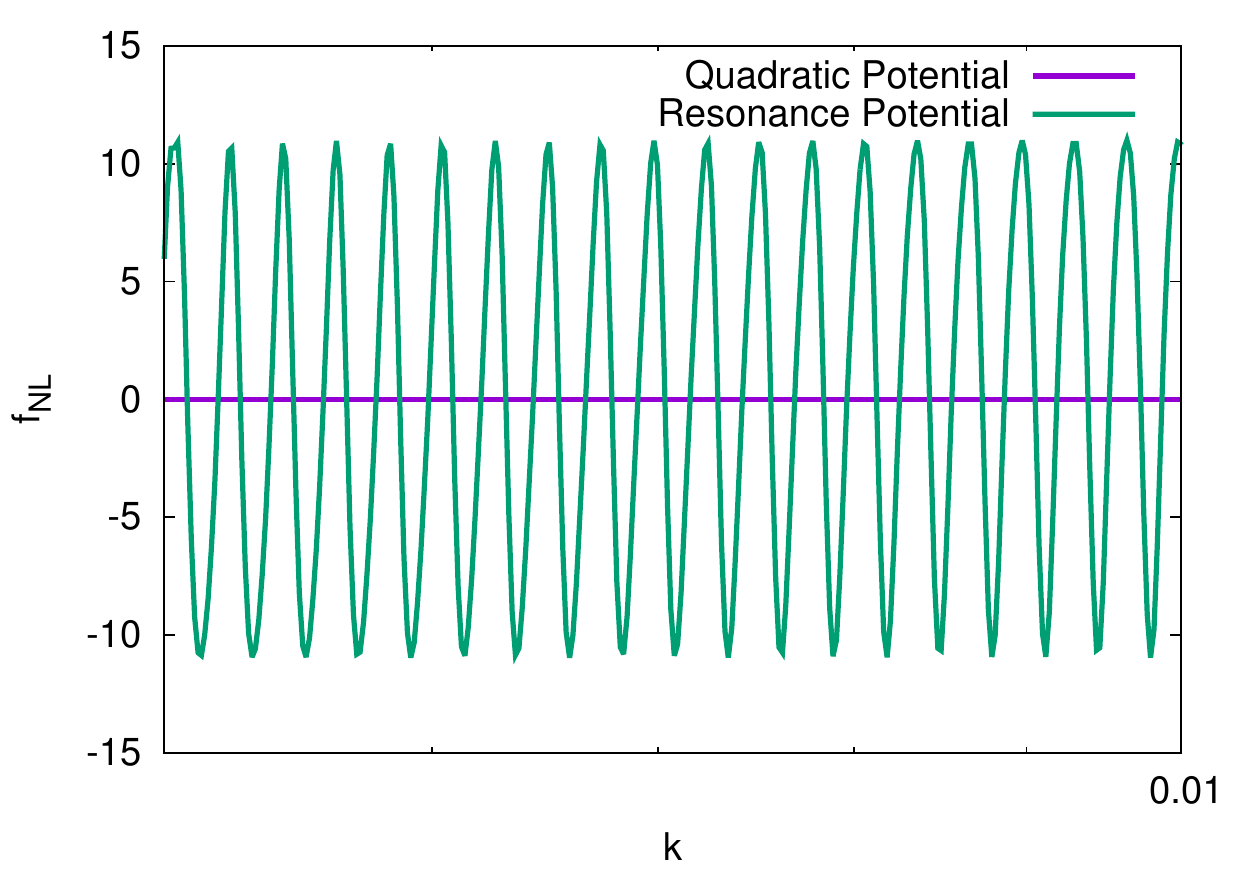}
\caption{$f_\textsc{NL}$ for resonance model plotted against $k=k_1=k_2=k_3$ for
equilateral triangle case for $c=5\times10^{-7}$ and $0.0007$.}
\label{fnlr}
\end{center}
\end{figure}
In Fig.~\ref{fnls} we present the calculation of
$f_\textsc{NL}$ for equilateral triangle for the step potential model with
parameters $c=0.002$, $d=0.02\;M_{pl}$ and $\phi_s=15.86\;M_{pl}$. The mode function $u_{k}$
that exits horizon at $\phi=\phi_s$ corresponds to $k_s\approx0.01\;M_{pl}$.  It
is noted that peaks in $f_\textsc{NL}$ plot for the step potential are of
$O(10)$ in amplitude that is about $500$ times the $f_\textsc{NL}$ of quadratic
potential. However, if we move away from the step $k_s$, $k>0.5\;M_{pl}$ and
$k<0.005\;M_{pl}$, the value of $f_\textsc{NL}$ for step potential tend to
approach values for the quadratic potential.

The other kind of models that can give rise to large non-Gaussianity are models
with global features in the potential like small oscillation on top of
quadratic potential also know as Resonance model\cite{XChen3}
\begin{eqnarray}
V_{res}(\phi)=\frac{1}{2}m^2\phi^2\left ( 1+ c \sin(\phi/\Lambda)  \right).
\label{step}
\end{eqnarray}
where $c$ is the amplitude of oscillation and
$\Lambda$ is the frequency. In this model the non-Gaussianity is generated in
the sub-horizon scales when modes are deep inside the horizon and they 
resonate, or interfere constructively, with the physical frequency
$\omega\approx\frac{\dot{\phi}}{2\pi\Lambda}$. This model introduces a new
scale $\Lambda$ into the quadratic potential as there small ripples in the
entire plane of the potential. These ripples in the potential causes
oscillation in $k$ space in the generalized $f_\textsc{NL}$ with amplitude
$\frac{5c/4}{\Lambda^{2.5}\phi^{0.5}}$(see Fig. \ref{fnlr}) while the power
spectrum remains almost flat with tiny ripples\cite{XChen3}. This kind of
mechanism may be realized in brane inflation with duality
cascade\cite{Bean,Girma}. Fig.~\ref{fnlr} shows our computational technique
works well for this kind of models as well.

%%%%%%%%%%%%%%%%%%%%%%%%%%%%%%%%%%%%%%%%%%%%%%%%%%%%%%%%%%%%%%%%%%%%%%%%%%%%%%%%%%%%%%%%%%%%%%%%%%%%%%%%%%%% Configuration Space Moments %%%%%%%%%%%%%%%%%%%%%%%%%%%%%%%%%%%%%%%%%%%%%%%%%%%%%%%%%%%%%%%%%%%%%%%%%%%%%%%%%%%%%%%%%%%%%%%%%%%%%%%%%%%%%%%%%%%%

\section{Configuration Space Moments}\label{MG}
Our goal is to develop a numerical technique for calculating 
the configuration space moments for perturbation $\zeta$ field in a general single-field model
of inflation. For that we need to integrate three-point function in momentum space as
given by Eqs.~(\ref{Mm1}-\ref{Mm3}). As a starting point we shall
consider slow roll models where these moments can be calculated analytically
for the case of flat power spectrum $P_k^{\zeta}\approx const$ 
and then use the acquired insight to develop a general numerical procedure. 

For instance, to calculate the $\left\langle \zeta^3(x) \right\rangle$,
we substitute Eq.~\ref{3A} into Eq.~\ref{Mm1} to get
\begin{eqnarray}
\left\langle \zeta^3 \right\rangle = 
\int  \prod_{i=1}^3 \frac{dk_i^3}{k_i^3}
\delta^3(\mathbf{k}_1+\mathbf{k}_2+\mathbf{k}_3)
\frac{(P_k^{\zeta})^2}{4\pi^2}
\mathcal{A}_{\mathbf{k}_1,\mathbf{k}_2, \mathbf{k}_3}
%e^{i(\bold{k}_1+\bold{k}_2+\bold{k}_3).\bold{x}}
%\\ \mathcal{A}=\frac{\eta^*-\epsilon^*}{8}\sum_i k_i^3 +
%\frac{\epsilon^*}{8}\sum_{i \ne j} k_ik_j^2 + \frac{\epsilon^*}{K}\sum_{i>j}
%k_i^2k_j^2\text{. } \label{z3}
\end{eqnarray}
where $\mathcal{A}$ is given by Eq.~\ref{A3p}. For near flat spectra
such integrals generally have both infrared and ultraviolet
divergences, thus the correspondent moment is formally infinite. Ultraviolet divergences
are regularized by the finite resolution of our measurements and
should be studied in the context of a specific experiment.
Infrared divergencies, on the other hand, seem of more conceptual nature 
since they come from  contribution of the modes much larger than the observed
Universe, which potentially reflect also less understood physics. 
To study them
we introduce infrared $k_{min}$ and ultraviolet $k_{max}$ cutoffs 
and consider asymptotic behaviour as $z=k_{min}/k_{max} \to 0$.
This procedure is somewhat ambiguous as to the order of imposing the cutoff
and integrating over the $\delta$-function. This ambiguity does not, however,
affect the coefficient of the leading infrared divergent term, which is, as we will see, 
what matters. It does affect subleading terms and this freedom can be used
to minimize them in the numerical calculations. We suggest the procedure
to use the $\delta$-function to eliminate in every term in the integral
\textit{the least infrared divergent momentum} and then integrate over the
angles and the magnitudes of remaining two momenta in 
$k_{min}$ and $k_{max}$ limits. 

Using the above mentioned procedure to evaluate the momentum space integrals,
we also calculated the other two moments
$\left\langle \zeta^2(x)\Delta\zeta(x) \right\rangle$,
$\left\langle (\nabla\zeta(x))^2\Delta\zeta(x) \right\rangle$ by substituting
Eq.~\ref{3A} in to Eqs. \ref{Mm2} and \ref{Mm3}. The asymptotic results
are given below up to $z^2$ order
\begin{widetext}
\begin{eqnarray}
\left\langle \zeta^3(x) \right\rangle &=&
(P_k^{\zeta})^2\left(\frac{3}{2} (\eta^* +2\epsilon^*) \ln^2(z)
-3\epsilon\ln (z)+ \epsilon\left(15-24
\ln(2)\right)+O(z^2)\right),\quad\label{z3}\\ 
\left\langle \zeta^2(x)\Delta\zeta(x) \right\rangle &=&-(P_k^{\zeta})^2 k_{max}^2 
\left((\eta^* + 2\epsilon^*)\ln(z) - \frac{\epsilon^*}{6}\left(\pi^2-28+32\ln(2)\right)+\mathcal{O}(z^2) \right),\label{q2z}\\ 
\left\langle(\nabla\zeta(x))^2\Delta\zeta(x) \right\rangle &=& 
-(P_k^{\zeta})^2k_{max}^4
\left(\frac{1}{6}\eta^*+\epsilon^*\left(\frac{4}{3}\ln(2)-\frac{527}{900}
\right)+O(z^2) \right).\label{q2I}
\end{eqnarray}
\end{widetext}
Evidently, the first two
moments are divergent in $z\rightarrow 0$ limit, 
$\left\langle \zeta^3(x) \right\rangle \propto \ln^2(z)$ while
$\left\langle\zeta^2(x)\Delta\zeta(x)\right\rangle \propto \ln(z)$.
The modified regularization sequence retains the
coefficients in the divergent terms but changes the constant ones.
Here we note that the last moment with the most derivatives 
$\left\langle (\nabla\zeta(x))^2\Delta\zeta(x) \right\rangle$ is
finite as $z \to 0$ and thus remains sensitive to regularization details.

%The above results are found by calculating the momentum space integrals by eliminating
%the least infrared divergent momentum in every term with help of
%$\delta(k_1+k_2+k_3)$ function and placing the cutoff on the remaining two
%momenta. But if we eliminate some other momenta with help of the delta function
%then we get different results from above expressions for the moments. However,
%the leading order terms, of order $\ln^2z$ and $\ln z$, in the first two
%moments come out to be the same but next to leading order terms are incorrect.
%While, the third moment that is finite is corrupted the most if we use the
%wrong momentum prescription. Thus, we eliminate the least divergent momenta in
%the infrared limit using the delta function.

Now, if we look at the expression for Euler characteristic
(Eq.~\ref{genus}) or any other observables, the configuration space  moments are always divided\cite{Pogosyan1}
by the variances
$\sigma^2$ and $\sigma_1^2$ given in Eqs.~\ref{sigmas}.
\begin{eqnarray}
\sigma^2 = \left\langle \zeta^2(x)
\right\rangle = P_k^{\zeta}\ln\left(\frac{k_{max}}{k_{min}}\right)\quad\\
\sigma_1^2 = \left\langle (\nabla\zeta(x))^2 \right \rangle
=P_k^{\zeta}\frac{k^2_{max}-k^2_{min}}{2}~.
\label{sigmas}
\end{eqnarray}
Thus, even though the moments by themselves are divergent, the observed
quantities are finite. Normalized or scaled by these variances the moments to
leading order in slow roll are given by
\begin{eqnarray}
S_3 &\equiv& \frac{\left\langle \zeta^3(x) \right\rangle}{\sigma^4}
\simeq\frac{3}{2}(\eta^* +2\epsilon^*),\quad\label{x3}\\
T_3 &\equiv& \frac{\left\langle \zeta^2(x)\Delta\zeta(x) \right\rangle}
{\sigma^2\sigma_1^2}\simeq-2(\eta^* +2 \epsilon^*),\label{q2x}\\
U_3 &\equiv& \frac{\left\langle (\nabla\zeta(x))^2\Delta\zeta(x) \right\rangle}{\sigma_1^4}\notag\\
&\simeq& -\frac{2}{3} \left(\eta^*+\epsilon^*\left(8\ln(2)-\frac{527}{150}
\right)\right).
\end{eqnarray}
Note that the numerical coefficient in front of $\epsilon^*$ in $U_3$ is also
equal to two with better than $2\%$ accuracy, thus all three normalized
moments are determined by $\eta^* + 2 \epsilon^*$. From the above moments, we have found 
that the local non-Gaussianity parameter for single-field slow roll inflation
is $f_\textsc{NL}=-\frac{5}{12}(\eta^*+ 2\epsilon^*)$ which is independent of the shape and scale of the triangles.

\begin{figure}
\begin{center}
\includegraphics[height=6.3cm]{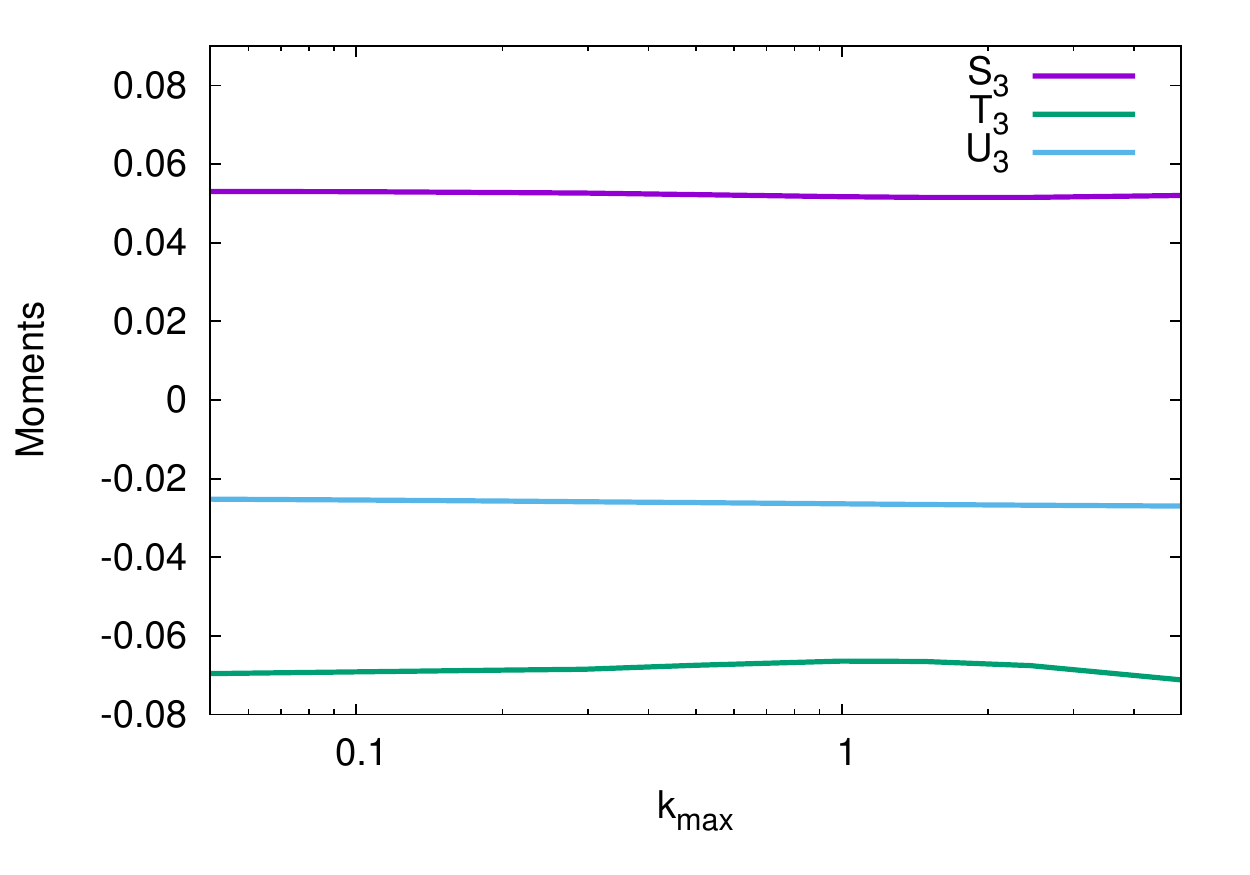}
\caption{Moments plotted against $k_{max}$ with fixed $k_{min}=0.005\;M_{pl}$
for quadratic potential.}\label{M0}
\end{center}
\end{figure}

For numerical calculation of momentum space integrals finite momentum space cutoff is
unavoidable. The above analytical results guide us to the following numerical procedure
for the general single-field model. We calculate the moments
using the exact result of 3-point function Eq.~\ref{3pt}, 
with the triangularity condition $\bold{k}_1+\bold{k}_2+\bold{k}_3=0$
applied to the least divergent momentum.To numerically obtain $z \to 0$ result
we integrate the remaining two momenta in finite range $[k_{min}$,$k_{max}]$,
then vary $\frac{k_{min}}{k_{max}}$ and find the asymptotic limit (plateau value) for the
scaled moments $S_3$, $T_3$ and $U_3$. Procedure we follow is that we fix
$k_{min}=0.005\;M_{pl}$ that corresponds to the mode that inflates $64$ e-folds
after horizon crossing that is roughly the largest scales of the observed
Universe and vary $k_{max}$ or $z$. As seen in Fig.~\ref{M0},
for quadratic potential the numerical calculation of 
$S_3$, $T_3$ and $U_3$ gives very stable result already by 
$k_{max}\approx 0.1\;M_{pl}$.
This fast convergence reflects smallness of the constant terms in 
Eqs.~\ref{z3} and \ref{q2z}, $-1.6 \epsilon$ and $0.8 \epsilon$ correspondingly.
Slight drop in values of moments in Fig.~\ref{M0} is a
numerical artifact as higher values of $k_{max}>2M_{pl}$ require very fine
resolution $\Delta k$ in momenta integrations. 

Thus we see that the infrared divergences for near flat spectra 
can be dealt with very efficiently during numerical momenta integration. 
It is sufficient to integrate over just a decade of wave-numbers to 
obtain a good approximation to the asymptotic values.  
The quadratic potential gives the values $S_3=0.0517$,
$T_3=-0.0664$ and $U_3=-0.0264$ with at most $5\%$ error in the
calculation of these moments.

\begin{figure}
\begin{center}
\includegraphics[height=6.3cm]{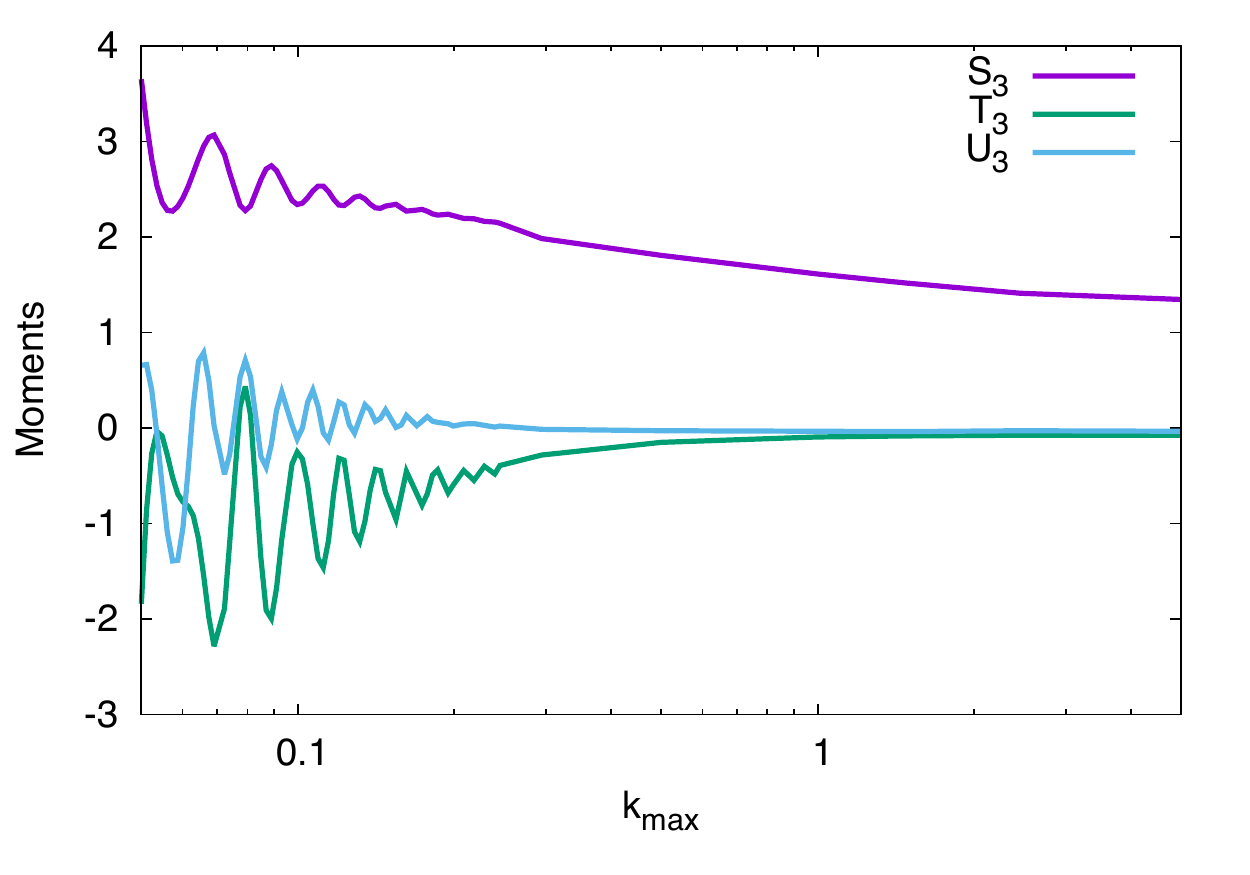}
\caption{Moments plotted against $k_{max}$ with
fixed $k_{min}=0.005\;M_{pl}$ for step potential with
$c=0.002$, $d=0.02\;M_{pl}$ and $\phi_s=15.86\;M_{pl}$.}
\label{Ms}
\end{center}
\end{figure}

For potential with features such as step potential, the answer for the moments
depends on the range of k-integration in relation 
to the modes that give additional contribution
to non-Gaussian signal. For the step potential, as Fig.~\ref{fnls} demonstrates,
the support for correction to the $f_\textsc{NL}$ is finite,
encompassing the range from $k \approx 0.01\;M_{pl}$ to $k\approx 0.2\;M_{pl}$
for the parameters used there. 
Fig.~\ref{Ms} shows the change in the moments if we fix $k_{min}=0.005\;M_{pl}$
but vary $k_{max}$.
The asymptotic behaviour can be understood from the following consideration.
Let us write
\begin{equation}
S_3^{step}(k_{min},k_{max})  = S_3^{quad} + 
\frac{\Delta \left\langle\zeta^3\right\rangle}{\sigma^4(k_{min},k_{max})}, 
\end{equation}
and correspondingly for the other moments, 
where $\Delta \left\langle\zeta^3\right\rangle$
signifies additional contribution to the momentum integral
above the baseline quadratic values. 
%Such split is useful if the additional contribution has finite $k$  support.
In such a split the first term, 
$S_3^{quad}$ is practically independent on the 
$k_{min}$ or $k_{max}$, as we discussed above. 
The second correction term depends on the integration range. As soon as this
range encompasses the support for the extra non-Gaussian contribution, it
starts decreasing as the correspondent powers of the variances,
i.e. as $1/\sigma^{4}\approx \ln^{-2} (k_{max}/k_{min})$ for $S_3$,
$1/(\sigma^{2}\sigma_1^{2}) \approx k_{max}^{-2} \ln^{-1}(k_{max}/k_{min})$ for $T_3$ and
$1/\sigma_1^{4} \approx k_{max}^{-4} $ for $U_3$. This explains the quick 
convergens of $T_3$ and $U_3$ moments to their baseline values as 
$k_{max} > 1\;M_{pl}$ and slow logarithimic decrease of $S_3$ exhibited
in Fig.~\ref{Ms}.

In cosmological applications the $k_{min}$ to $k_{max}$ range depends on
observational setup or analysis choice. For large-scale structure studies
the ratio of largest observable scale $\sim 10\;Gpc$ to the smallest
presently mildly non-linear scale $\sim 10\;Mpc$ is of order of
a thousand, which points to $k_{max} \sim 5\;M_{pl}$.
This is the value that we adopt in the following sample calculations.

\begin{figure}[htbp]
\begin{center}
\includegraphics[height=6.3cm]{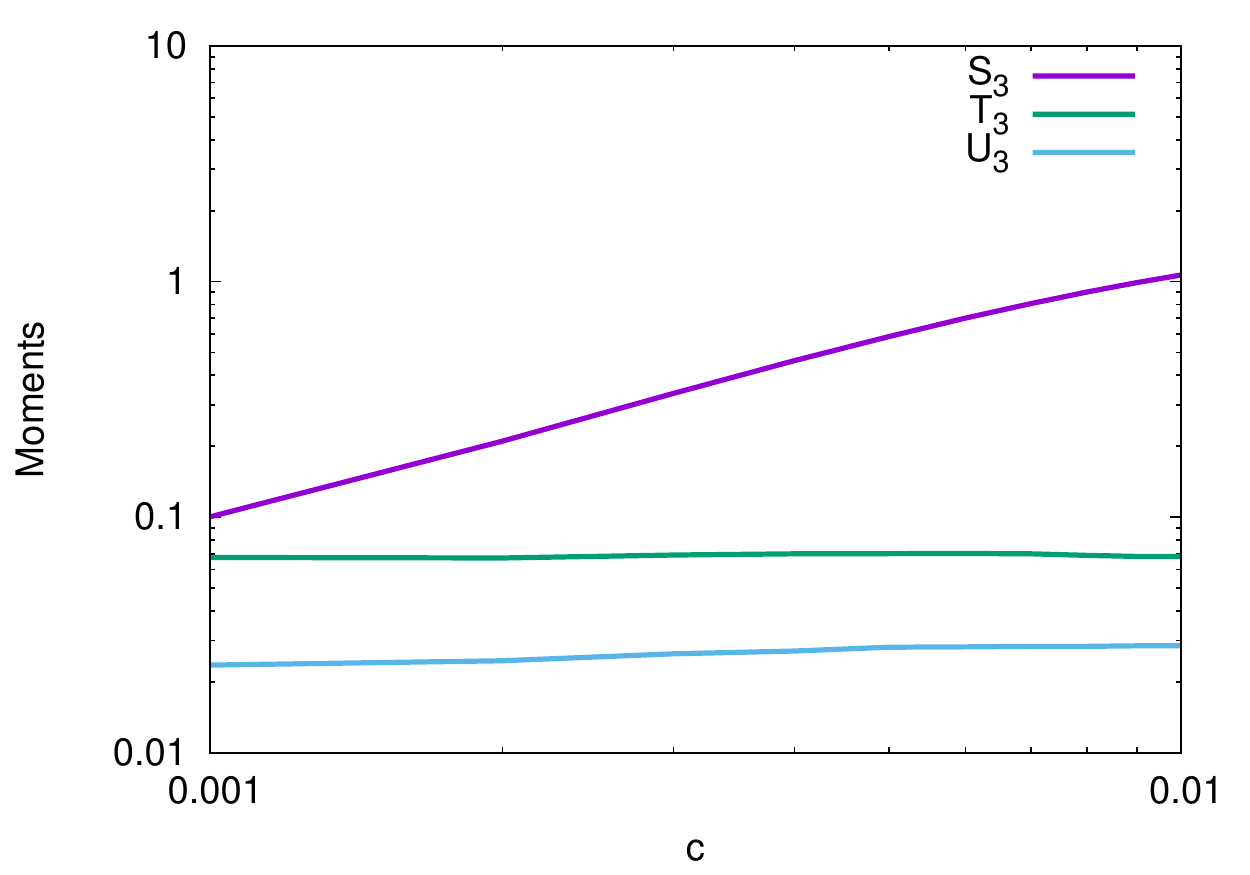}
\caption{Absolute values of the moments for the step potential plotted against parameter $c$ for fixed $d=0.01\;M_{pl}$.} 
\label{Mc}
\end{center}
\end{figure}

\begin{figure}[htbp]
\begin{center}
\includegraphics[height=6.3cm]{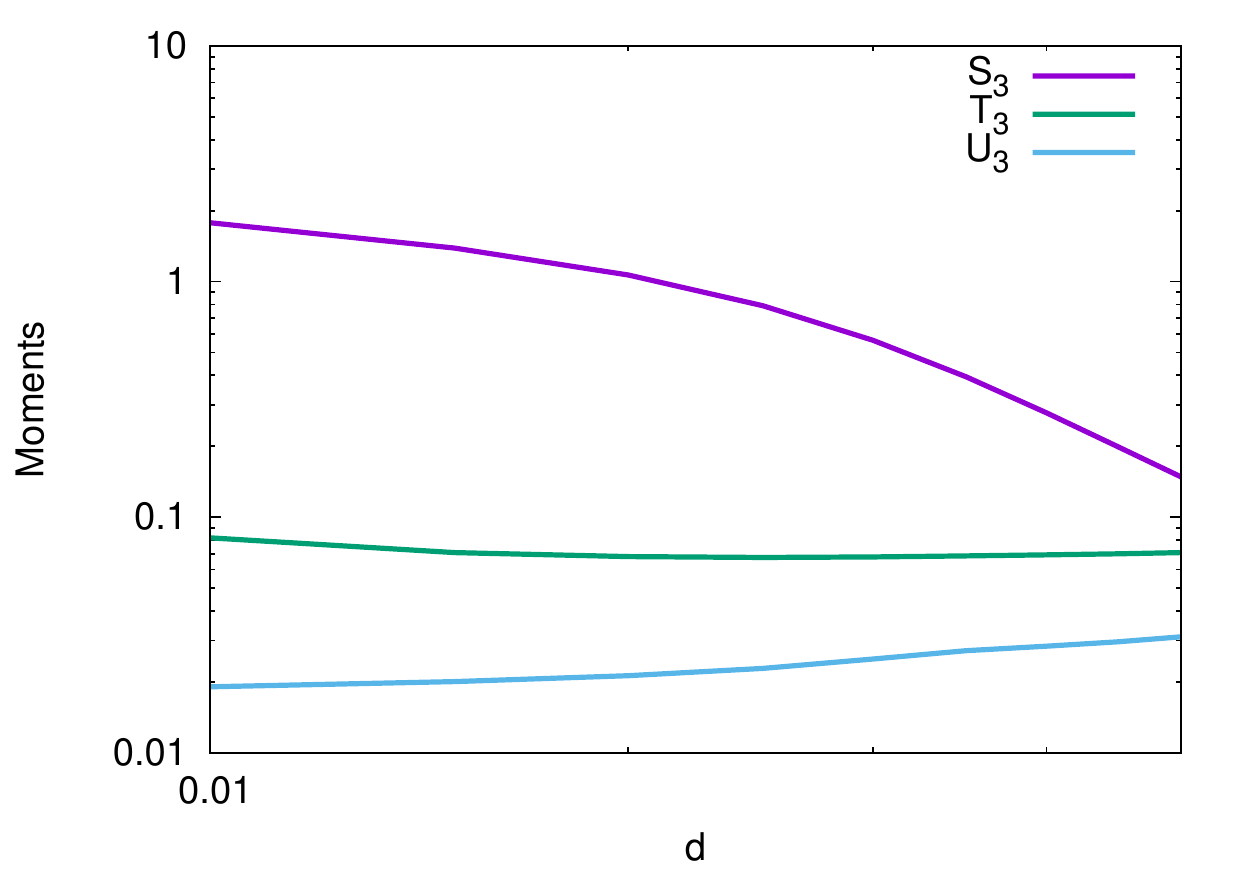}
\caption{Absolute values of the moments for the step potential plotted against parameter $d$ for fixed $c=0.01$.}
\label{Md}
\end{center}
\end{figure}

\begin{figure}[htbp]
\begin{center}
\includegraphics[height=6.1cm]{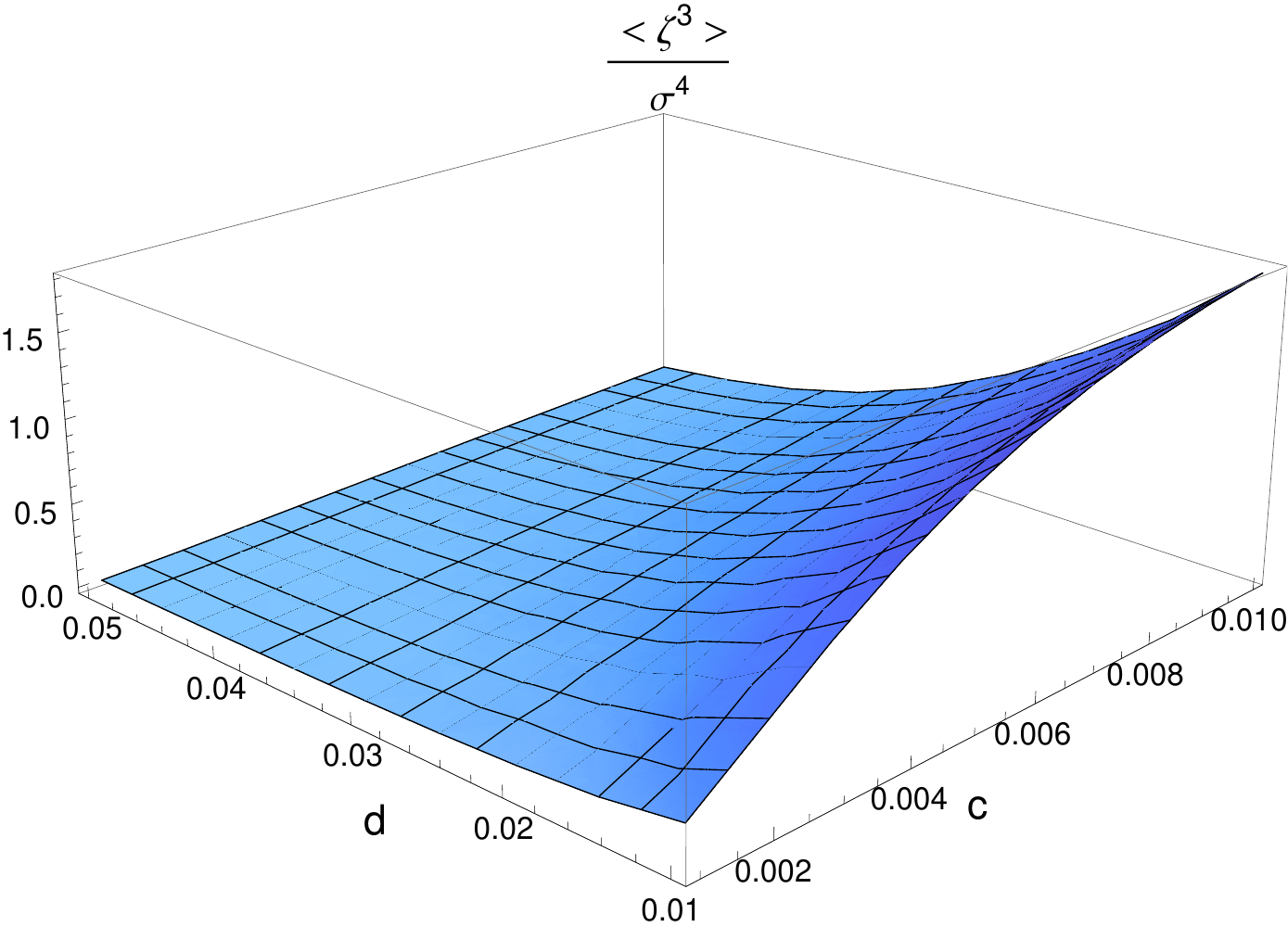}
\caption{Moment $S_3=\frac{\left\langle \zeta^3\right\rangle}{\sigma^4}$ for the step potential
plotted against parameters $c$ and $d$ showing that this moment peeks when the step is sharpest.}
\label{Mcd}
\end{center}
\end{figure}

For the step potential we plot the moments
against the parameters $c$ and $d$ as displayed  in Figs.~\ref{Mc} and
\ref{Md}. The limit of the quadratic potential is reached as $c \to 0$
or $d \to \infty$.
It can be seen that the magnitude of
$S_3$ moment increases significantly as the height of the step $c$ increases
(Fig~\ref{Mc}) or the width $d$ decreases (Fig.~\ref{Md}).
At the same time, the moments $T_3$ and $U_3$ remain practically unchanged
from their small values in quadratic potential limit, as expected
from our analysis of $k_{max}$ behaviour.
The first moment $S_3$ is plotted against $c$ and $d$ in
Fig.~\ref{Mcd} which shows that sharper the step, smaller width and higher
step, the higher the moment.

%%%%%%%%%%%%%%%%%%%%%%%%%%%%%%%%%%%%%%%%%%%%%%%%%%%%%%%%%%%%%%%%%%%%%%%%%%%%%%%%%%%%%%%%%%%%%%%%%%%%%%%%% Geometrical Statistics and Minkowski Functionals %%%%%%%%%%%%%%%%%%%%%%%%%%%%%%%%%%%%%%%%%%%%%%%%%%%%%%%%%%%%%%%%%%%%%%%%%%%%%%%%%%%%%%%%%%%%%%%%%%%%%%%%%%%%%%

\section{Geometrical Statistics and Minkowski Functionals} 
Many geometrical statistics, including Minkowski functionals, of the mildly
non-Gaussian fields can be expressed as series expansion in higher-order
moments of the field and its derivatives with powers of $\sigma$ controlling
the expansion order \cite{PGP09}. First order non-Gaussian corrections 
are linear in $\sigma$ and defined by cubic normalized moments that
we have studied earlier.  For the models with
nearly flat power spectrum that we consider, we adopt the value of the variance 
$\sigma \approx \sqrt{P_\zeta \ln(k_{max}/k_{min})} \approx 1.2 \times 10^{-4}$ with $P_\zeta = 2.2 \times 10^{-9}$ suggested by Planck data \cite{planck2}
and $k_{max}/k_{min} = 10^3$.

\begin{figure}[htbp]
\begin{center}
\includegraphics[height=5.2cm]{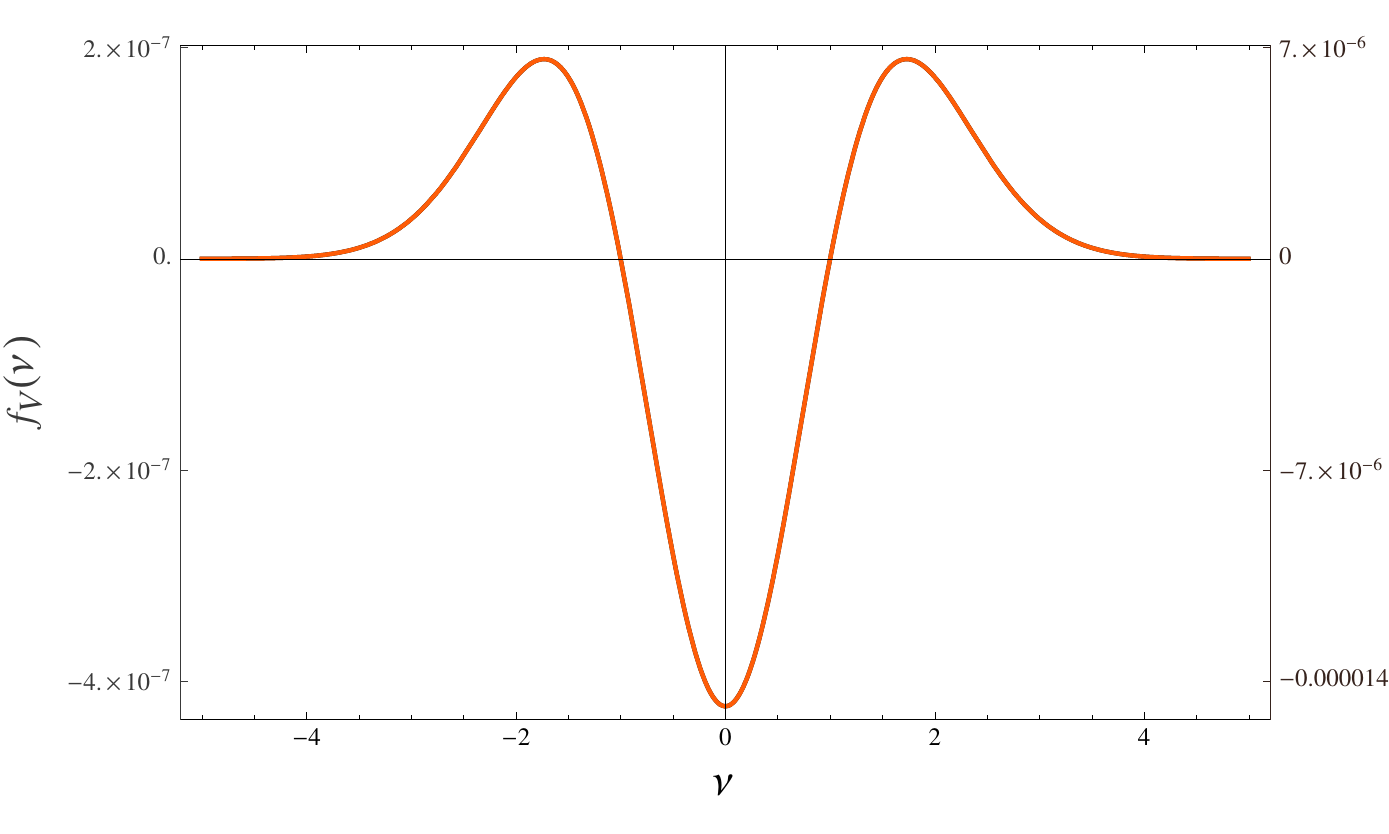}
\caption{Non gaussian part of filling factor $f_V$ as the function of 
threshold $\nu$. Values on the left vertical axis are for quadratic potential 
while values on the right axis are for step potential with $c=0.01$ and $d=0.01\;M_{pl}$.}
\label{f0}
\end{center}
\end{figure}

The simplest geometrical statistic or Minkowski functional is the filling
factor $f_V=\int_{\nu\sigma}^\infty d \zeta P(\zeta)$,
i.e. the fraction of volume above the threshold $\nu$. 
Its moment expansion gives
\begin{eqnarray}
f_V(\nu)=\frac{1}{2}\mathrm{Erfc}\left( \frac{\nu}{\sqrt{2}} \right)
+\frac{e^{-\frac{\nu^2}{2}}}{\sqrt{2\pi}}\sigma H_2(\nu)\frac{S_3}{6} +O(\sigma^2).
\label{fnu}
\end{eqnarray}
The first non-Gaussian 
correction only depends on $S_3$ multiplied by Hermite polynomial of order two.
Thus, linear in $\sigma$ non-Gaussian part of $f_V(\nu)$ 
has a global shape that is independent of any model
while its magnitude will depend on the magnitude of moment $S_3$.
Thus for the step potential, the filling factor can be as large as $500$
times the $f_V$ for quadratic potential.
In Fig. \ref{f0} a single curve shows the non-Gaussian part of the filling
factor both  for the model with quadratic potential with $f_V$ correction
of order $\mathcal{O}(10^{-7})$ (as labeled on the left vertical axis) and 
for the step potential for which it is $\mathcal{O}(10^{-5})$ 
(the right vertical axis).

There are several advantages to use the value of the filling factor $f_V$
instead of $\nu$ as a variable in which to express all other statistics.
Indeed, the fraction of volume occupied by a data set is often available
even from the limited data, whereas specifying $\nu$ requires prior knowledge 
of the variance $\sigma$ which may not be easily obtainable.
In some cases non-Gaussian analysis itself gives more robust way to determine
the variance. Following \cite{Matsubara2} we introduce the effective threshold
$\nu_f \equiv \sqrt{2} \mathrm{Erfc}^{-1}(2 f_V)$ be used as an observable
alternative to $\nu$. To the first order correction in $\sigma$ we have the relation
\begin{equation}
\nu = \nu_f + \sigma \frac{S_3}{6} H_2(\nu_f).
\end{equation}

\begin{figure}[htbp]
\begin{center}
\includegraphics[height=5.2cm]{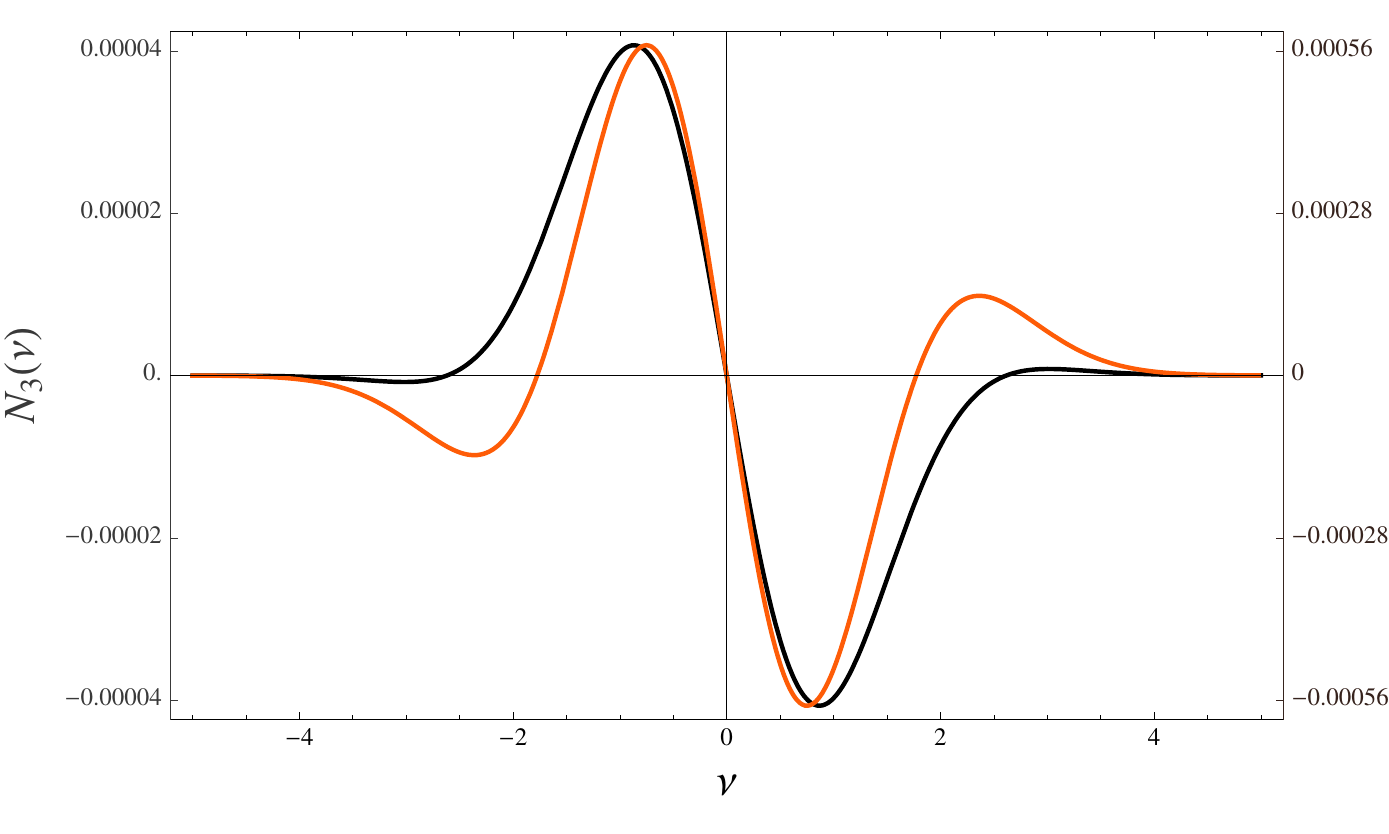}
\caption{Non gaussian part of area of isodensity contours $\mathcal{N}_3(\nu)$ for quadratic potential in black with values on left vertical axis 
as a function of threshold $\nu$. Similarly, $\mathcal{N}_3(\nu)$ for step potential, for $c=0.01$ and $d=0.01\;M_{pl}$, is show in red(lighter colour) with values on right vertical axis as a function of threshold $\nu$.}
\label{N1}
\end{center}
\end{figure}

Another Minkowski functional in 3D is the area (per unit volume) of isodensity contours $\mathcal{N}_3(\nu)$. 
Up to first non-Gaussian contribution, this quantity is expressed in terms of third-order moments as
\begin{eqnarray}
\mathcal{N}_3(\nu) \approx\frac{2e^{-\frac{\nu^2}{2}}\sigma_1}{\sqrt{3}\pi\sigma}\Bigg[1+\sigma\left(\frac{S_3}{6}H_3(\nu) + \frac{T_3}{2}H_1(\nu)\right)  \Bigg],~\label{N3v}
\end{eqnarray}
depending on moments $S_3$ and $T_3$.
In Fig.~\ref{N1} we show the non-Gaussian $\mathcal{N}_3(\nu)$ 
for quadratic and step potential. Besides vastly different amplitudes,
it exhibits different shape that distinguishes the two models. For quadratic potential 
$H_3$ contribution is small and $\mathcal{N}_3$ has just two measurable extrema
with small secondary ones near the edges, while
due to the large value of $S_3$ the $H_3$ part is prominent for the step
potential, giving rise to four distinct extrema, as seen in Fig.~\ref{N1}.

\begin{figure}[htbp]
\begin{center}
\includegraphics[height=5.2cm]{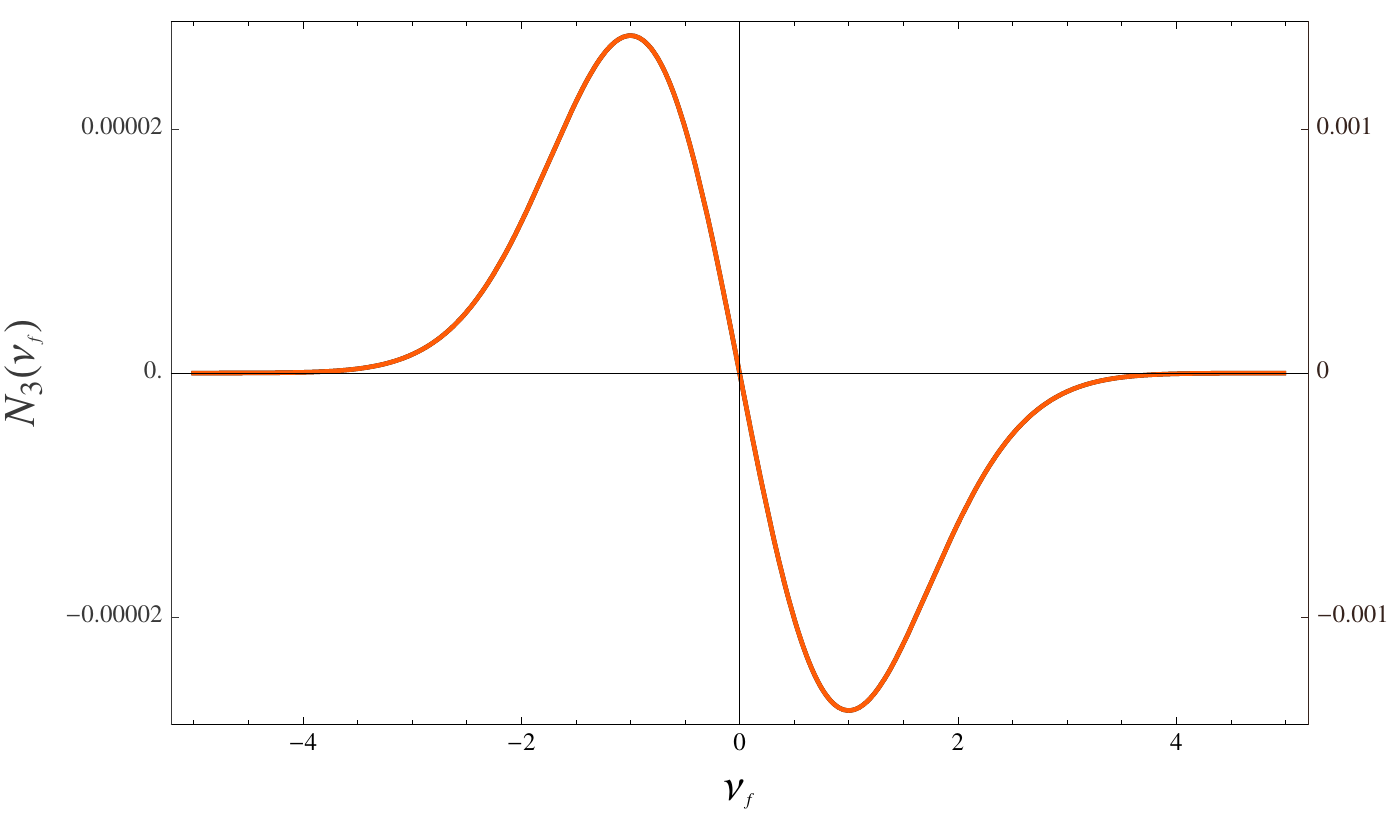}
\caption{Non gaussian part of area of isodensity contours $\mathcal{N}_3(\nu_f)$ for quadratic potential in black with values on left vertical axis 
as a function of filling factor $\nu_f$. Similarly, $\mathcal{N}_3(\nu_f)$ for step potential, for $c=0.01$ and $d=0.01\;M_{pl}$, in red(lighter colour) with values on right vertical axis as a function of filling factor $\nu_f$.}
\label{N2}
\end{center}
\end{figure}

As a function of $\nu_f$, i.e of the filling factor, 
the area of isodensity contours is
\begin{eqnarray}
\mathcal{N}_3(\nu_f) \approx\frac{2e^{-\frac{\nu_f^2}{2}}\sigma_1}{\sqrt{3}\pi\sigma}\Bigg[1-\sigma\left(S_3+\frac{T_3}{4}\right)H_1(\nu_f)  \Bigg],\quad\label{N3f}
\end{eqnarray}
which demonstrates the general outcome of eliminating the highest order Hermite
polynomial term when switching from $\nu$ to $\nu_f$.
Because of that, in Fig.~\ref{N2}, the area of isodensity contours $\mathcal{N}_3(\nu_f)$ has a model independent shape that is proportional to $e^{-\frac{\nu_f^2}{2}}H_1(\nu_f)$. The $\mathcal{N}_3(\nu_f)$ for the step potential has values of order $10^{-3}$ on right axis while for quadratic potential it has values of order $10^{-5}$ on the left axis.

An important Minkowski functional in cosmology is the Euler characteristic
(or genus) that is used to characterize the topology of the isocontours of random fields. The
definition generally used by cosmologists is that the genus is the number of
holes minus the number of isolated regions above a threshold $\nu$ in a random
field (thus, for one isolated region, the genus is just the number of holes minus one),
while Euler characteristic of the excursion set is just minus genus, see Eq.~\ref{genus}.
Euler characteristic density can be also considered as a function of
filling-factor-deduced threshold $\nu_f$ rather than the threshold $\nu$.
\begin{eqnarray}
%{\scriptscriptstyle 3D}
\chi_\textsc{3D}(\nu_f) \approx \frac{1}{(2\pi)^{2}}\left(\frac{\sigma_1}{\sqrt{3}
\sigma} \right)^3  e^{-\nu_f^2/2}\Bigg[ H_2(\nu_f) - \Bigg.\notag \\
\Bigg.\sigma\left( S_3 + \frac{3}{4}T_3 \right)H_3(\nu_f) -\sigma\left(S_3 +
\frac{9}{4}U_3 \right)H_1(\nu_f)   \Bigg]\quad
\label{genus_f}
\end{eqnarray}

%The above expression for genus have Hermite polynomial of order $3$ and $1$
%that are multiplied by moments $S_3$, $T_3$ and $U_3$. 

\begin{figure}[htbp]
\begin{center}
\includegraphics[height=5.2cm]{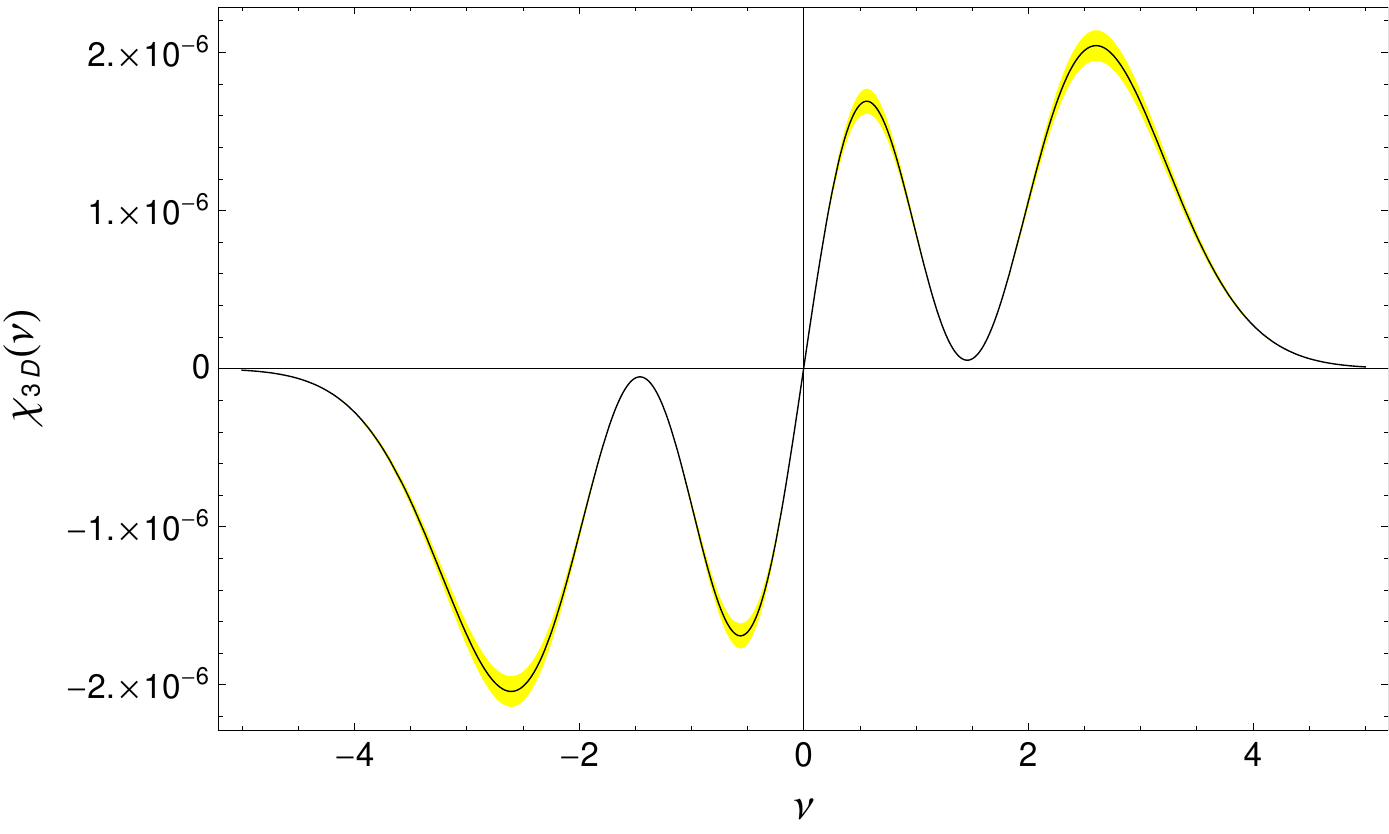}
\caption{Non gaussian part of Euler characteristic $\chi_\textsc{3D}$ as a function of threshold $\nu$ for the quadratic potential. The
shaded region shows the error bars coming from the error in calculation of moments that are roughly $5$ percent.}
\label{G0v}
\end{center}
\end{figure}

\begin{figure}[htbp]
\begin{center}
\includegraphics[height=5.2cm]{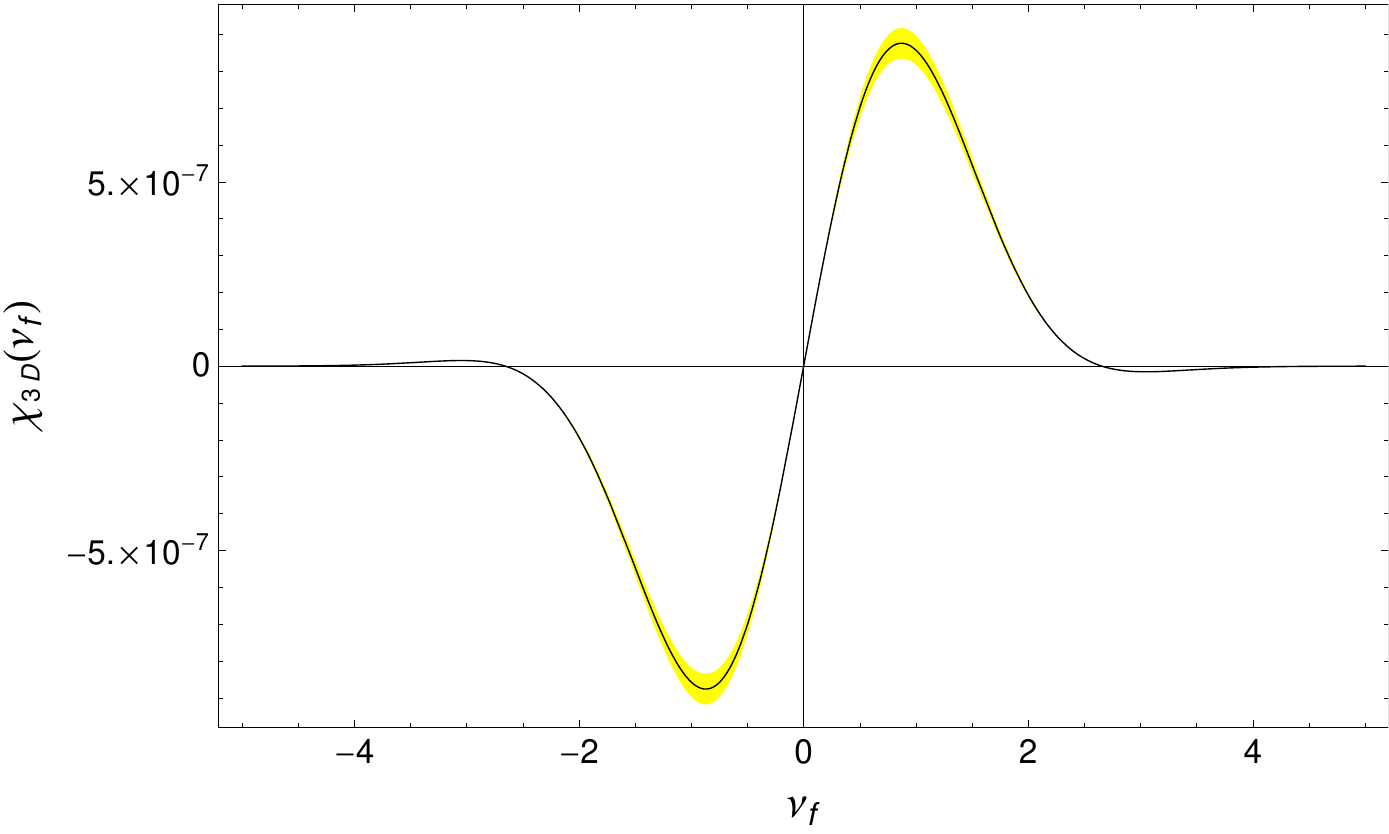}
\caption{Non gaussian part of genus as a function of filling factor $\nu_f$ for the quadratic potential. The
shaded region shows the error bars coming from the error in calculation of moments that are roughly $5$ percent.}
\label{G0f}
\end{center}
\end{figure}

We have calculated Euler characteristic $\chi_\textsc{3D}$ of the 3D perturbation field
$\zeta$ for two models of inflation, the quadratic inflation and step potential model. 
For quadratic potential, the non-Gaussian part of $\chi_\textsc{3D}$ as a function of threshold $\nu$ is plotted in Fig.~\ref{G0v}.
showing the small amplitude of the non-Gaussianity of order $10^{-7}$ and
the presence of all three $H_1, H_3, H_5$ harmonics.
Euler characteristic as a function of $\nu_f$ is given in Fig.~\ref{G0f}.  
For the quadratic inflation $S_3+3T_3/4 \approx 0.002$ is notably smaller
than $S_3+9 U_3/4 \approx -0.008$, hence the result is dominated by the $H_1$
term that has only one zero crossing at origin as can be seen in Fig.~\ref{G0f}.

\begin{figure}[htbp]
\begin{center}
\includegraphics[height=5.2cm]{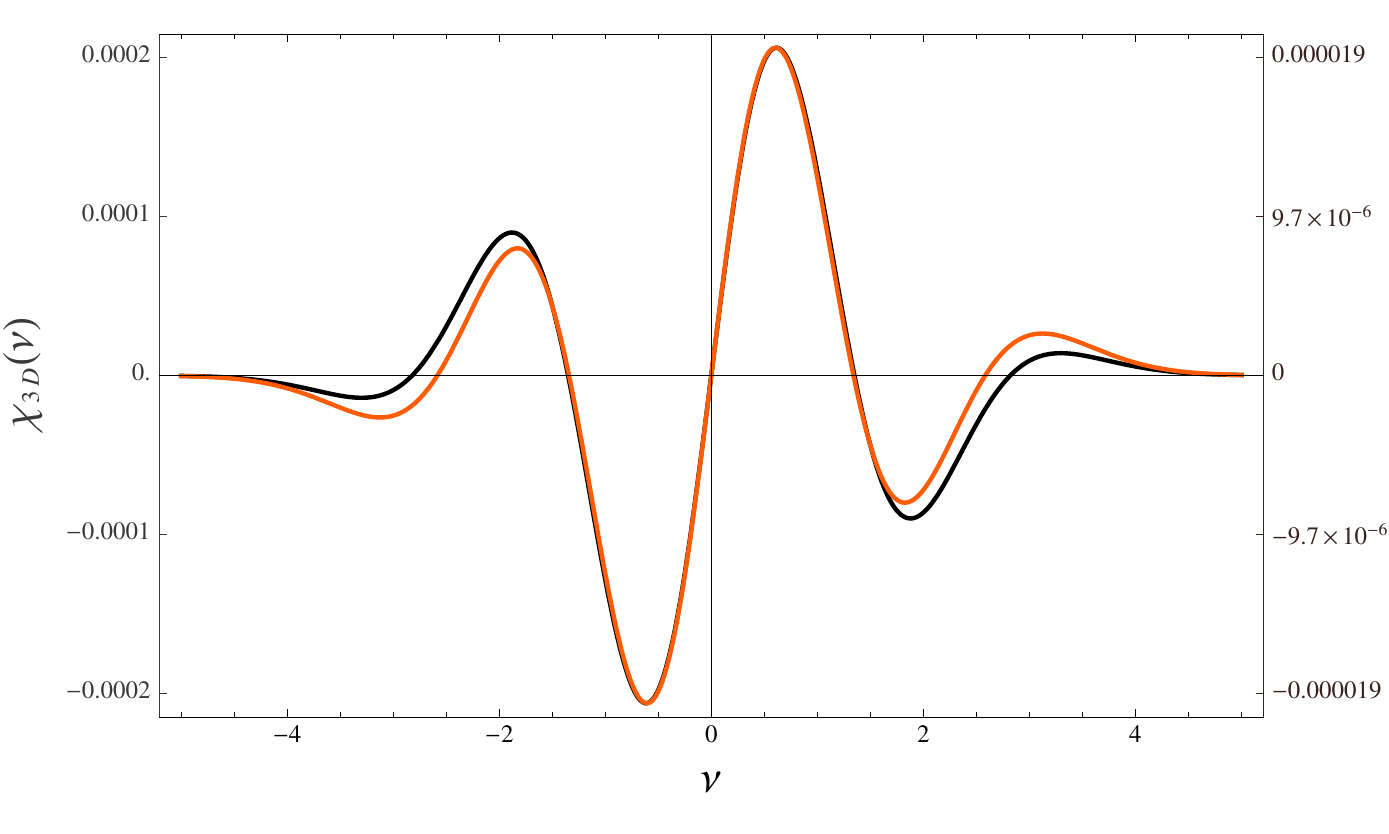}
\caption{Non gaussian part of genus as a function of threshold $\nu$ for the step potential for two parameter sets
($c=0.0018$, $d=0.022\;M_{pl}$ in black on left axis) and ($c=0.01$, $d=0.01\;M_{pl}$ in red, lighter colour, on right axis).}
\label{G1}
\end{center}
\end{figure}

\begin{figure}[htbp]
\begin{center}
\includegraphics[height=5.2cm]{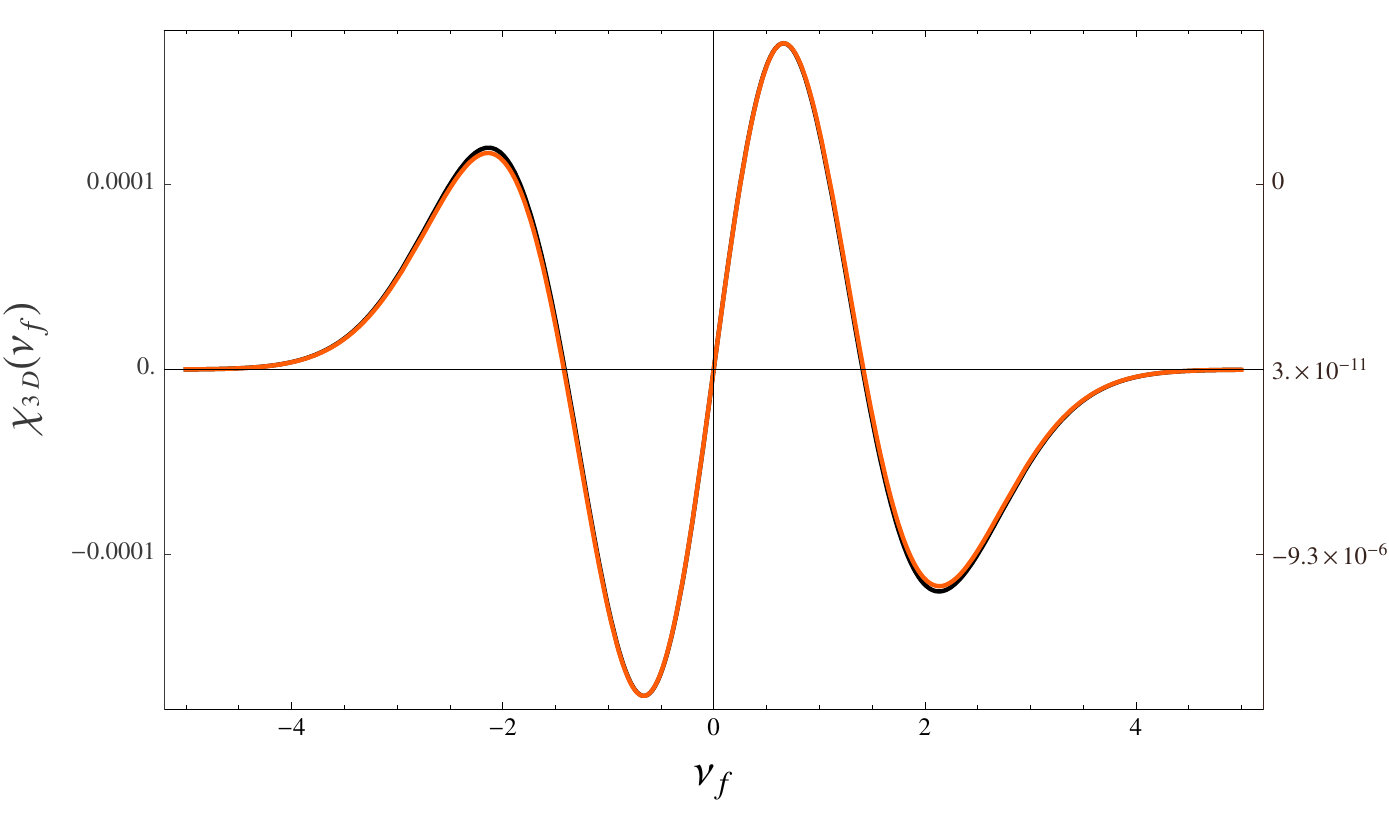}
\caption{Non gaussian part of genus as a function of filling factor $\nu_f$ for the step potential for two parameter sets
($c=0.0018$, $d=0.022\;M_{pl}$ in black on left axis) and ($c=0.01$, $d=0.01\;M_{pl}$ in red, lighter colour, on right axis).}
\label{G2}
\end{center}
\end{figure}

While the $\mathcal{O}(10^{-7})$ non-Gaussian correction to Euler characteristic for the
quadratic potentialis small, as expected, and is hardly observable, 
for the step potential, it will have the magnitude $10$ to $500$ times larger. 
Fig.~\ref{G1} shows $\chi_\textsc{3D}(\nu)$ curves for two 
different sets of $c$ and $d$ parameters, with magnitude of the effect 
differing by an order of magnitude, namely
$\mathcal{O}(10^{-5})$ for $c=0.0018$ and $d=0.022\;M_{pl}$ and
$\mathcal{O}(10^{-4})$ for $c=0.01$ and $d=0.01\;M_{pl}$.
At the same time the shapes of the Euler characteristic curves are very similar dominated by $S_3 H_5(\nu)$ term.
This can be seen explicitly also when measuring Euler characteristic as
a function of $\nu_f$ as shown in Fig.~\ref{G2}. 
With $S_3$ dominant for all the values of step parameters $c$ and $d$ studied,
the non-Gaussian part of Euler characteristic exhibits a universal form
$\chi_\textsc{3D} \propto \sigma S_3 \left(H_1(\nu_f) + H_3(\nu_f) \right)$
that is a distinguishing signature of this particular model of non-Gaussianity.

%%%%%%%%%%%%%%%%%%%%%%%%%%%%%%%%%%%%%%%%%%%%%%%%%%%%%%%%%%%%%%%%%%%%%%%%%%%%%%%%%%%%%%%%%%%%%%%%%%%%%%%%%%%% Results and Discussion %%%%%%%%%%%%%%%%%%%%%%%%%%%%%%%%%%%%%%%%%%%%%%%%%%%%%%%%%%%%%%%%%%%%%%%%%%%%%%%%%%%%%%%%%%%%%%%%%%%%%%%%%%%%%%%%%%%%%%%%

\section{Results and Discussion}
The main goal in this paper was to develop theoretical formalism
that links early Universe inflationary models
to the observable geometrical characteristics
of the initial field of scalar adiabatic cosmological perturbations.
The link to non-Gaussian features in such statistics as Minkowski functionals,
extrema counts, and skeleton properties is provided by studying the
higher order moments of the perturbation field and its derivatives in
configuration space.

We have investigated from the first principles the third-order configuration
space moments that give first non-Gaussian corrections.
To calculate these moments we have to calculate the
three-point bispectrum in momentum space and then integrate over the three
momenta with appropriate combinations of derivative operators.

We presented a complete prescription for the calculation of three-point
function, in momentum space, for general single-field models of inflation,
improving on the computational procedures used in the previous literature. We
have devised a method to precisely calculate the required time integral over
products of perturbation mode functions
in ``before horizon crossing'' as well as in ``after horizon crossing'' regimes.
For that we stressed the necessity to use the action that gives stable three-point function integral
result till
the very end of inflation \cite{Arroja} and we developed novel technique
using modified Cesaro summation
to integrate the highly oscillatory
``before horizon crossing'' part of the three-point function integral. Thus, we
calculated the bispectrum in terms of generalized $f_\textsc{NL}$ that even
works for models that break slow roll conditions and where there is significant
variation in slow roll parameters.

In the configuration space we have studied infrared and ultraviolet divergences
of the third-order perturbation moments. Analytical analysis of the
slow-roll models allowed us to illuminate the role of the dominant
infrared divergent logarithms in the moment calculations. These terms determine
observable moments scaled by the appropriate powers of the variance of the
field or its derivative.  The moments that do not contain infrared divergence
are found to be sensitive to ultraviolet smoothing prescription.
We implemented this understanding in numerical procedure that transforms
bispectrum results to the configuration space moments.

To calculate the moments we had to integrate out triangles of all shapes and
sizes. We introduced the regularization in the momentum space integrals
with both infrared $k_{min}$ and ultraviolet $k_{max}$ cuts. We
observed that for the basic flat spectrum 
these moments do not depend on $k_{min}$,
instead leading logarithmic terms depend on the ratio
$z=\frac{k_{min}}{k_{max}}$ of the smallest the largest scales
and can be robustly select relative to subdominant term by 
varying this ratio which is numerically easier to do by varying $k_{max}$.
It was further noted that these moments divided by
their corresponding variances are finite quantities and they reach a plateau
for large values of $k_{max}$ hence they are also independent of physics of the
smallest scales (see Figs. \ref{M0} and \ref{Ms}). 
For the step potential it was shown that the moment
$S_3$ is the one that is affected the most  while the
other two moments $T_3$ and $U_3$ are very mildly affected
by the step if the range of $k$ integration encompasses all the modes
with enhanced non-Gaussian response.

%In the step
%potential, the oscillatory behaviour of the generalized $f_\textsc{NL}$ or
%bispectrum, gives a mild contribution to the moments because the momenta
%integration are dominated by large momenta while the step is located in the
%small momentum limit. 

The Minkowski functionals of the cosmological perturbation field generated
on inflation depend on the balance between the moments of the field and
its derivatives up to the second order. We have demonstrated typical behaviour
of the non-Gaussian corrections to several Minkowski functionals of
$\zeta$ field in the slow-roll quadratic model and the 
model with step potential.  
Such corrections show different signatures for different models of inflation.
In particular step potential, as representative of the class of models with
enhanced non-Gaussian response over limited wavelength range, 
leads to a characteristic universal form of the genus curve as
function of the filling factor if measurements include a sufficiently
large interval of scales.

\section{Acknoledgements}
This research has been supported by Natural Sciences 
and Engineering Research Council (NSERC) of Canada 
Discovery Grant. The computations were performed on the 
SciNet HPC Consortium. SciNet is funded by: the Canada 
Foundation for Innovation under the auspices of Compute 
Canada; the Government of Ontario; Ontario Research 
Fund - Research Excellence; and the University of Toronto\cite{SciNet}.

\bibliographystyle{unsrt} \bibliography{ThesisRefs}

\end{document}